\documentclass[twocolumn]{revtex4}
\pdfoutput=1
\usepackage{graphicx}
\usepackage{dcolumn}
\usepackage{longtable}
\usepackage{amsmath}
\usepackage{amssymb}

\newcommand{\tbeta}{\tilde{\beta}}

\begin{document}

\title
{Bohr Hamiltonian with deformation-dependent mass term
for the Davidson potential}

\author
{Dennis Bonatsos$^1$, P. E. Georgoudis$^1$, D. Lenis$^1$, N. Minkov$^2$, and C. Quesne$^3$}

\affiliation
{$^1$Institute of Nuclear Physics, National Centre for Scientific Research 
``Demokritos'', GR-15310 Aghia Paraskevi, Attiki, Greece}

\affiliation
{$^2$Institute of Nuclear Research and Nuclear Energy, Bulgarian Academy of Sciences, 72 Tzarigrad Road, 1784 Sofia, Bulgaria}

\affiliation
{$^3$ Physique Nucl\'eaire Th\'eorique et Physique Math\'ematique, Universit\'e Libre de Bruxelles, Campus de la Plaine CP229,
Boulevard du Triomphe, B-1050 Brussels, Belgium}

\begin{abstract}

Analytical expressions for spectra and wave functions are derived for a Bohr Hamiltonian, 
describing the collective motion of deformed nuclei, in which the mass is allowed to depend on the nuclear deformation. 
Solutions are obtained for separable potentials consisting of a Davidson potential 
in the $\beta$ variable, in the cases of $\gamma$-unstable nuclei, axially symmetric prolate deformed 
nuclei, and triaxial nuclei, implementing the usual approximations in each case. 
The solution, called the Deformation Dependent Mass (DDM) Davidson model, 
is achieved by using techniques of supersymmetric quantum mechanics (SUSYQM),
involving a deformed shape invariance condition. Spectra and $B(E2)$ transition rates are compared to experimental data. 
The dependence of the mass on the deformation, dictated by SUSYQM for the potential used, 
reduces the rate of increase of the moment of inertia with deformation, removing a main drawback of the model. 

\end{abstract}


\maketitle
    
\section{Introduction}

The Bohr Hamiltonian \cite{Bohr} and its extensions, the geometrical 
collective model \cite{BM,EG}, have provided for several decades a sound 
framework for understanding the collective behaviour of atomic nuclei. 
It has been customary to consider in the Bohr Hamiltonian the mass 
to be a constant. However, evidence has been accumulating that this approximation 
might be inadequate. In particular:

1) The moments of inertia are predicted to increase proportionally to $\beta^2$, 
where $\beta$ is the collective variable corresponding to nuclear deformation, 
while the experimentally determined (from the spectra) moment of inertia 
shows a much more moderate increase as a function of the experimentally determined 
(from the $B(E2)$ transition rates) deformation, especially for well deformed 
nuclei \cite{Ring}. This discrepancy has led to arguments that the use of the 
Bohr Hamiltonian is justified  for vibrational and transitional nuclei,
but its applicability to deformed nuclei needs further clarification. 

2) Detailed comparisons to experimental data have recently pointed out \cite{Jolos1,Jolos2} 
that the mass tensor of the collective Hamiltonian cannot be considered as
a constant and should be taken as a function of the collective coordinates, with 
quadrupole and hexadecapole terms present in addition to the monopole one.  

3) In the framework of the Interacting Boson Model (IBM) \cite{IA}, which offers 
an algebraic description of atomic nuclei complementary to that of the Bohr Hamiltonian, 
it is known that in its geometrical limit \cite{IA}, obtained through the use 
of coherent states \cite{IA}, terms of the form $\beta^2 \pi^2$ and/or more complicated terms appear \cite{vanRoos},
in addition to the usual term of the kinetic energy, $\pi^2$. Thus it might
be appropriate to search for a modified form of the Bohr Hamiltonian, in
which the kinetic energy term will be modified by terms containing $\beta^2$ and/or more complicated terms. 

Based on this evidence, a Bohr Hamiltonian with a mass depending on the collective variable $\beta$ 
can be considered. Position-dependent effective masses have been studied recently 
in a general framework \cite{QT4267}, while several Hamiltonians known to be soluble 
through techniques of supersymmetric quantum mechanics (SUSYQM) \cite{PR,SUSYQM}, 
have been appropriately generalized \cite{Q2929} to include position-dependent effective masses, 
the 3-dimensional harmonic oscillator being among them \cite{Q2929}. 

In the present work we are going to show that a Bohr Hamiltonian with a Davidson potential \cite{Dav} in $\beta$ 
(a harmonic oscillator potential with a term proportional to $1/\beta^2$ added to it)
can be generalized in order to include a mass depending on $\beta$, $B=B_0/(1+a\beta^2)^2$, where $B_0$ and $a$ are constants.
We shall call this approach the Deformation Dependent Mass (DDM) Davidson model. 
Three cases of potentials, for which exact separation of variables can be achieved, will be considered: 

\noindent
a)  Potentials independent \cite{Wilets} of the collective variable $\gamma$ (an angle measuring departure from axial symmetry), 
called $\gamma$-unstable potentials, appropriate for describing vibrational and near-vibrational nuclei.
 
\noindent
b) Potentials of the form \cite{Wilets,F1,F2,F3,ESDPRC}
$ v(\beta,\gamma)= u(\beta)+ w(\gamma)/\beta^2$, with $u(\beta)$ being the Davidson potential \cite{Dav}, 
and with $w(\gamma)$ having a deep minimum at $\gamma=0$, corresponding to axially symmetric prolate deformed nuclei.

\noindent
c) Potentials of the form
$ v(\beta,\gamma)= u(\beta)+ w(\gamma)/\beta^2$, with $u(\beta)$ being the Davidson potential \cite{Dav}, 
and with $w(\gamma)$ having a deep minimum at $\gamma=\pi/6$, corresponding to triaxial nuclei \cite{DF,DR}.

Analytical results for spectra and $B(E2)$ transition rates will be provided for all three cases,
implementing the usual approximations in each limit \cite{IacX5,MtV,Z5}, 
while comparison to experimental results will be undertaken  in the first two, for which able bulk 
of experimental data exists. A special solution regarding $\gamma$-unstable nuclei has been given earlier in 
Ref. \cite{first}.  

The analytical spectra and wave functions of the Bohr Hamiltonians considered are obtained by using techniques 
of supersymmetric quantum mechanics \cite{PR,SUSYQM}, equivalent \cite{SUSYQM} to the factorization method of Infeld and Hull
\cite{Infeld}. The integrability of the Hamiltonian is achieved by imposing a deformed shape invariance condition 
\cite{Q2929}. These tools are described in more detail in Section~VI.  

It should be noticed that the concept of a non-constant mass in the framework of the Bohr Hamiltonian 
has been used long ago in numerical solutions of a generalized Bohr Hamiltonian \cite{Kumar}, as well as 
in relevant mean field calculations \cite{Libert}. The main difference of the present work from these earlier 
approaches is that analytical solutions are obtained here. In addition, in the present case the number of free parameters remains small
(two or three), while the functional dependence of the mass on the deformation for the potential used is dictated by SUSYQM. 
The relation of the present work to these earlier approaches will be discussed in Section XII.  

The structure of the present work is as follows. In Section II the formalism of position-dependent effective masses,
which we use in order to allow the mass to depend on the deformation $\beta$, is briefly reviewed, and applied 
to the Bohr Hamiltonian in Section  III. The three exactly separable cases described above are considered in Section IV,
in which the common overall form of the radial equation in all three cases is pointed out, while in Section V 
we focus on the use of the Davidson potential in the radial equation. The solvability of the Hamiltonian is achieved 
in Section VI by imposing a deformed shape invariance condition, leading to the energy spectrum given in Section VII
and the wave functions given in Section VIII. Normalization coefficients are given  in Section IX, while a detail 
on their numerical calculation is included as Appendix 1. $B(E2)$ transition probabilities are considered in Section X, 
while in Section XI comparisons of spectra and $B(E2)$s to experimental data are carried out. Finally, connections 
to earlier work are discussed in Section XII, while Section XIII
contains discussion of the present results and plans for further work.  

\section{Formalism of position-dependent effective masses} 

For reasons of completeness, we briefly review the basics of the formalism 
needed in handling effective masses depending on the position. 
The main problem encountered is the generalization of the kinetic energy term.
We show how this can be solved in an unambiguous way. 

When the mass $m({\bf x})$ is position dependent \cite{QT4267}, 
it does not commute 
with the momentum ${\bf p} = -i\hbar \nabla$. Therefore, there 
are many ways to generalize the usual form of the kinetic energy, 
${\bf p}^2 /(2 m_0)$, where $m_0$ is a constant mass, in order to obtain 
a Hermitian operator. In order to avoid any specific choices, one can use
the general two-parameter form proposed by von Roos \cite{vRoos},
with a Hamiltonian 
\begin{eqnarray}
H = -{\hbar^2\over 4} [m^{\delta'}({\bf x}) \nabla m^{\kappa'}({\bf x})
\nabla m^{\lambda'}({\bf x}) \nonumber \\
+ m^{\lambda'}({\bf x}) \nabla m^{\kappa'}({\bf x}) \nabla m^{\delta'}({\bf x})]+ V({\bf x}),
\end{eqnarray}     
where $V$ is the relevant potential and the parameters $\delta'$, $\kappa'$,
$\lambda'$ are constrained by the condition $\delta'+ \kappa' + \lambda'=-1$.
Assuming  a position dependent mass of the  form 
\begin{equation}
m({\bf x}) = m_0 M({\bf x}), \quad M({\bf x})=\frac{1}{(f({\bf x}))^2}, 
\quad f({\bf x})=  1+ g ({\bf x}), 
\end{equation}
where $m_0$ is a constant mass and $M({\bf x})$ is a dimensionless 
position-dependent mass, the Hamiltonian becomes 
\begin{eqnarray}\label{Hdelta}
H = -{\hbar^2 \over 4 m_0} [f^{\delta}({\bf x}) \nabla f^{\kappa}({\bf x})
\nabla f^{\lambda}({\bf x}) \nonumber \\+ f^{\lambda}({\bf x}) \nabla f^{\kappa}({\bf x})
\nabla f^{\delta}({\bf x}) ]  +V({\bf x}),
\end{eqnarray}
with $\delta + \kappa +\lambda=2$. It is known \cite{QT4267} that this 
Hamiltonian can be put into the form 
\begin{equation}\label{eq:e3}
H =-{\hbar^2 \over 2 m_0} \sqrt{f({\bf x})} \nabla f({\bf x}) \nabla 
\sqrt{f({\bf x})} +V_{eff}({\bf x}), 
\end{equation} 
with
\begin{eqnarray}
V_{eff}({\bf x}) = V({\bf x}) +{\hbar^2 \over  2 m_0} \left[ {1\over 2} 
(1-\delta-\lambda) f({\bf x}) \nabla^2 f({\bf x}) \right. \nonumber \\
+\left. \left({1\over 2}-\delta
\right) \left( {1\over 2}-\lambda\right) (\nabla f({\bf x}))^2 \right],  
\end{eqnarray}
where $\delta$ and $\lambda$ are free parameters.

In the final part of the paper, in which comparison to experiment
will be carried out by fitting the theoretical predictions to the experimental data, 
it will be seen that the predictions for the theoretical spectra 
turn out to be independent of the choice made for $\delta$ and $\lambda$. 

\section{Bohr Hamiltonian with deformation-dependent effective mass} 

\subsection{Deformation-dependent effective mass formalism} 

The original Bohr Hamiltonian \cite{Bohr} is
\begin{eqnarray}\label{eq:e1}
H_B = -{\hbar^2 \over 2B} \left[ {1\over \beta^4} {\partial \over \partial 
\beta} \beta^4 {\partial \over \partial \beta} + {1\over \beta^2 \sin 
3\gamma} {\partial \over \partial \gamma} \sin 3 \gamma {\partial \over 
\partial \gamma} \right. \nonumber \\
\left. - {1\over 4 \beta^2} \sum_{k=1,2,3} {Q_k^2 \over \sin^2 
\left(\gamma - {2\over 3} \pi k\right) } \right] +V(\beta,\gamma),
\end{eqnarray}
where $\beta$ and $\gamma$ are the usual collective coordinates
($\beta$ being a deformation coordinate measuring departure from spherical shape, 
and $\gamma$ being an angle measuring departure from axial symmetry), while
$Q_k$ ($k=1$, 2, 3) are the components of angular momentum in the intrinsic 
frame, and $B$ is the mass parameter, which is usually considered constant.  

We wish to construct a Bohr equation with a mass depending on
the deformation coordinate $\beta$,
in accordance with the formalism described above,  
\begin{equation}
B(\beta)=\frac{B_0}{(f(\beta))^2}, 
\end{equation}
where $B_0$ is a constant. We then need the usual Pauli--Podolsky prescription
\cite{Podolsky} 
\begin{equation}\label{Pod}
(\nabla \Phi)^i = g^{ij} {\partial \Phi \over \partial x^j}, \qquad 
\nabla^2 \Phi = {1\over \sqrt{g}} \partial_i \sqrt{g} g^{ij} \partial_j \Phi,
\end{equation}
in order to construct a Schr\"{o}dinger equation corresponding to the Hamiltonian of
Eq. (\ref{eq:e3}) 
in a 5-dimensional space equipped with the Bohr-Wheeler
coordinates $\beta,\gamma$. Since the deformation function
$f$ depends only on the radial coordinate $\beta$, 
only the $\beta$ part of the resulting equation will be 
affected, the final result reading
\begin{equation}
\begin{split}
\label{eq:mBohr}
 & H \Psi = \left[ 
-{1\over 2} {\sqrt{f}\over \beta^4} {\partial \over \partial \beta} 
\beta^4 f {\partial \over \partial \beta} \sqrt{f}
-{f^2 \over 2 \beta^2 \sin 3\gamma} {\partial \over \partial \gamma} 
\sin 3\gamma {\partial \over \partial \gamma} \right.  \\
 & \left. + {f^2\over 8 \beta^2} 
\sum_{k=1,2,3} {Q_k^2 \over \sin^2\left(\gamma -{2\over 3} \pi k \right)}
+ v_{eff} \right] \Psi = \epsilon \Psi,
\end{split}  
\end{equation}
where reduced energies $\epsilon = B_0 E/\hbar^2$ and reduced potentials
$v= B_0 V/\hbar^2$ have been used, with
\begin{eqnarray}
v_{eff}= v(\beta,\gamma)+ {1\over 4 } (1-\delta-\lambda) f \nabla^2 f  \nonumber \\
+ {1\over 2} \left({1\over 2} -\delta\right) \left( {1\over 2} -\lambda\right)
(\nabla f)^2 .
\end{eqnarray}

\subsection{Connection to curved space} 

In Ref. \cite{QT4267} it has been proved that the position-dependent effective mass 
formalism can be equivalently expressed in a curved space. We shall prove here 
that this connection is possible also in the case of the Bohr Hamiltonian,
paving the way for connecting in Section XII the present results to earlier related work. 

Ordering the coordinates as 
\begin{equation}
q_1=\Phi, \quad q_2=\Theta, \quad q_3=\psi, \quad q_4=\beta, \quad q_5=\gamma, 
\end{equation}
the kinetic energy in the standard Bohr Hamiltonian \cite{Bohr} 
can be represented as 
\begin{equation}
T= {B\over 2} \left( ds \over dt \right)^2, 
\end{equation} 
where
\begin{equation}
ds^2= g_{ij} dq_i dq_j,
\end{equation}
the symmetric matrix $g_{ij}$ having the form 
\begin{equation}\label{gmatr}
(g_{ij})= \left( \begin{matrix} g_{11} & g_{12} & g_{13} & 0 & 0 \cr
                        g_{21} & g_{22} &   0    & 0 & 0 \cr
                        g_{31} &   0    & g_{33} & 0 & 0 \cr
                          0    &   0    &   0    & g_{44} & 0 \cr 
                          0    &   0    &   0    & 0  & g_{55} \cr \end{matrix}\right),  
\end{equation}
with \cite{Sitenko}
\begin{equation}\label{gmatrel}
\begin{split}
 & g_{11}= {{\cal J}_1 \over B} \sin^2\Theta \cos^2\psi +  {{\cal J}_2\over B} \sin^2\Theta \sin^2\psi
+ {{\cal J}_3\over B} \cos^2\Theta,  \\
 & g_{12}= {1\over B} ({\cal J}_2-{\cal J}_1) \sin\Theta \sin\psi \cos\psi ,  \\
 & g_{13}= { {\cal J}_3 \over B} \cos\Theta,  \\
 & g_{22}= {{\cal J}_1 \over B} \sin^2\psi + {{\cal J}_2 \over B} \cos
 ^2\psi,  \\
 & g_{33}= {{\cal J}_3 \over B},  \\
 & g_{44}=1 , \\
 & g_{55}=\beta^2,
\end{split}   
\end{equation}
where the moments of inertia are 
\begin{equation}
{\cal J}_k = 4 B \beta^2 \sin^2\left( \gamma - k {2\pi \over 3}  \right). 
\end{equation}
The determinant of the matrix is 
\begin{equation}\label{det}
g = { {\cal J}_1 {\cal J}_2 {\cal J}_3 \over B^3} \beta^2 \sin^2\Theta 
= 4 \beta^8 \sin^2 3\gamma \sin^2\Theta. 
\end{equation}
The relevant volume element is then 
\begin{equation}
dV = 2 \beta^4 \sin 3\gamma \sin\Theta d\Phi d\Theta d\psi d\beta d\gamma. 
\end{equation}
The inverse matrix is found to be 
\begin{equation}\label{igmatr}
(g^{-1}_{ij})= \left( \begin{matrix} g_{11}^{-1} & g_{12}^{-1} & g_{13}^{-1} & 0 & 0 \cr
                        g_{21}^{-1} & g_{22}^{-1} &  g_{23}^{-1}    & 0 & 0 \cr
                        g_{31}^{-1} &  g_{32}^{-1}    & g_{33}^{-1} & 0 & 0 \cr
                          0    &   0    &   0    & g_{44}^{-1} & 0 \cr 
                          0    &   0    &   0    & 0  & g_{55}^{-1} \cr \end{matrix}\right),  
\end{equation}
with 
\begin{equation}
\begin{split}
 & g_{11}^{-1} = {B\over \sin^2\Theta} \left({\cos^2 \psi \over {\cal J}_1} + {\sin^2\psi \over {\cal J}_2}  \right), \\
 & g_{12}^{-1} = -B \left({1\over {\cal J}_1}-{1\over {\cal J}_2}  \right)  {\sin\psi \cos\psi \over \sin\Theta}, \\
 & g_{13}^{-1}= -B \left({\cos^2 \psi \over {\cal J}_1} + {\sin^2\psi \over {\cal J}_2}  \right) {\cot\Theta \over \sin\Theta},  \\
 & g_{22}^{-1}= B \left({\sin^2 \psi \over {\cal J}_1} + {\cos^2\psi \over {\cal J}_2}  \right),  \\
 & g_{23}^{-1}= B \left({1\over {\cal J}_1}-{1\over {\cal J}_2}  \right) \cot\Theta \sin\psi \cos\psi ,  \\
 & g_{33}^{-1}= B \left({\cos^2 \psi \over {\cal J}_1} + {\sin^2\psi \over {\cal J}_2}  \right) \cot^2\Theta + {B\over {\cal J}_3},  \\
 & g_{44}^{-1}= 1, \\
 & g_{55}^{-1}= {1\over \beta^2}.
\end{split}
\end{equation} 
Using these matrix elements and the value of the determinant from Eq. (\ref{det}) in Eq. (\ref{Pod}) we obtain 
\begin{equation}
\begin{split}
& T= -{\hbar^2 \over 2 B} \nabla^2 =-{\hbar^2 \over 2B} \left[ {1\over \beta^4} {\partial \over \partial 
\beta} \beta^4 {\partial \over \partial \beta} \right. \\ 
& \left. + {1\over \beta^2 \sin 
3\gamma} {\partial \over \partial \gamma} \sin 3 \gamma {\partial \over 
\partial \gamma}
 - {1\over 4 \beta^2} \sum_{k=1,2,3} {Q_k^2 \over \sin^2 
\left(\gamma - {2\over 3} \pi k\right) } \right], 
\end{split}
\end{equation}
where $Q_k$ are the components of the angular momentum in the intrinsic frame
\begin{equation}
\begin{split}
 & Q_x= -i\left( -{\cos \psi \over \sin\Theta} {\partial \over \partial \Phi} + \sin \psi  {\partial \over \partial \Theta}
+ \cot \Theta \cos \psi {\partial \over \partial \psi} \right), \\
 & Q_y= -i\left( -{\sin \psi \over \sin\Theta} {\partial \over \partial \Phi} + \cos \psi  {\partial \over \partial \Theta}
- \cot \Theta \sin \psi {\partial \over \partial \psi} \right), \\
 & Q_z= -i {\partial \over \partial \psi}. 
\end{split}
\end{equation}

The connection between the position-dependent effective mass and curved spaces 
has been considered in Ref. \cite{QT4267}.  According to the findings of Ref. \cite{QT4267}, 
one expects in the present case all elements of the matrix (\ref{gmatr}) to be divided by $f^2$
\begin{equation}
g'_{ij}= {g_{ij}\over f^2}.
\end{equation}
As a result, the determinant of the matrix will be 
\begin{equation}
g' = {g \over f^{10}}, 
\end{equation}
and the volume element will be 
\begin{equation}
dV' = {dV\over f^5}. 
\end{equation}
The elements of the inverse matrix will be 
\begin{equation}
{g'}_{ij}^{-1} = f^2 g_{ij}^{-1}. 
\end{equation}

According to Ref. \cite{QT4267}, in order to obtain the Schr\"{o}dinger equation 
in the form of Eq. (\ref{eq:mBohr}), one has to start with the equation 
\begin{equation}\label{Hg}
H_g \tilde \Psi = \left[ -{1\over 2} \nabla^2 + u_g \right] \tilde \Psi
= \left[ -{1\over 2} {1\over \sqrt{g'}} \partial_i \sqrt{g'} {g'}_{ij}^{-1} \partial_j + u_g \right] \tilde \Psi,
\end{equation}
where
\begin{equation}
\tilde \Psi = f^{5/2} \Psi,
\end{equation}
while reduced energies and reduced potentials are used, as in Eq. (\ref{eq:mBohr}). 
The exponent in the last equation is related to the dimensionality of the space. 

Substituting the $g'$ matrix elements and determinant in Eq. (\ref{Hg}), 
and performing the relevant calculation (which closely resembles the pure Bohr case, 
except for the 44-term), 
we see that Eqs. (\ref{Hg}) and (\ref{eq:mBohr}) do coincide with 
\begin{equation} 
u_g = u_{eff}+ f \ddot f -2 (\dot f)^2 +4 {f \dot f \over \beta}, \quad \dot f= {df \over d\beta}, 
\quad \ddot f = {d^2 f \over d \beta^2}.
\end{equation}

This result has several important consequences.

1) It becomes clear that solving the Schr\"{o}dinger equation (\ref{eq:mBohr}) with deformation dependent mass
is equivalent to solving a modified Bohr equation (\ref{Hg}) with different metric matrix $g'$ and 
another effective potential, $u_g$. Between the two equivalent schemes, one chooses to solve Eq. (\ref{eq:mBohr})
instead of Eq. (\ref{Hg}), just because the former can be solved analytically through the use of SUSYQM techniques.

2) The wave functions $\tilde \Psi = f^{5/2} \Psi$ are accompanied by the volume element $dV'= dV/ f^5$.
As a result
\begin{equation}
\int \tilde \Psi^* \tilde \Psi dV' = \int  (f^{5/2} \Psi^*) (f^{5/2} \Psi) {dV \over f^5} = \int \Psi^* \Psi dV, 
\end{equation}
i.e., the wave functions $\Psi$ of the deformation dependent mass problem correspond to the usual Bohr volume element $dV$. 

3) The simple relation between $\tilde \Psi$ and $\Psi$ also shows that the wave functions $\Psi$ satisfy the well-known 
24 symmetries of Bohr wave functions \cite{Bohr}, which the wave functions $\tilde \Psi$ satisfy by construction.  
If these symmetries were not satisfied, the solutions could not have been used for the description of nuclei. 

Further consequences, regarding the connection of the present approach to earlier work, 
will be discussed in Section XII.

\section{Exactly separable special forms of the Bohr Hamiltonian} 

The solution of the above Bohr-like equation can be reached for
certain classes of potentials using techniques developed in the
context of SUSYQM \cite{PR,SUSYQM,Q2929}. 
At this point exact separation of variables can be achieved in three cases,
described in the following three subsections. 

\subsection{$\gamma$-unstable nuclei} 

In order to achieve separation of variables
we assume that the potential $v(\beta,\gamma)$ depends only 
on the variable $\beta$, i.e. $v(\beta)=u(\beta)$ \cite{Wilets}. 
Potentials of this kind are called $\gamma$-unstable potentials, since they are appropriate 
for the description of nuclei which can depart from axial symmetry without any energy cost. 

One then seeks wave functions of the form \cite{Wilets,IacE5}
\begin{equation}
\Psi (\beta, \gamma, \theta_i)= \xi(\beta) \Phi(\gamma, \theta_i),
\end{equation}
where $\theta_i$ ($i=1$, 2, 3) are the Euler angles.
Separation of variables gives
\begin{multline}\label{eq:e6}
\left[ 
-{1\over 2} {\sqrt{f}\over \beta^4} {\partial \over \partial \beta} 
\beta^4 f {\partial \over \partial \beta} \sqrt{f}
+{f^2\over 2\beta^2}  \Lambda 
+{1\over 4} (1-\delta-\lambda) f\nabla^2 f \right.\\ \left.  
+{1\over 2} \left( {1\over 2}-\delta\right) \left( {1\over 2} -\lambda\right)
(\nabla f)^2 +u(\beta) \right] \xi(\beta) = \epsilon \xi(\beta),  
\end{multline}
\begin{eqnarray}\label{eq:e7}
\left[ -{1\over \sin 3\gamma} {\partial \over \partial \gamma} \sin 3\gamma
{\partial \over \partial \gamma} + {1\over 4} \sum_k {Q_k^2 \over 
\sin^2\left( \gamma -{2\over 3} \pi k\right) } \right] \nonumber \\ 
\Phi (\gamma, \theta_i) = \Lambda \Phi(\gamma, \theta_i). 
\end{eqnarray}
Eq. (\ref{eq:e7}) has been solved by B\`es \cite{Bes}. 
$ \Lambda=\tau(\tau+3)$
represents the eigenvalues of the second order Casimir operator of SO(5), 
while $\tau$ is the seniority quantum number, characterizing the irreducible
representations of SO(5). The values of angular momentum $L$ occurring 
for each $\tau$ are provided by a well known algorithm and are listed in \cite{IA,Wilets}.
Within the ground state band (gsb) one has $L=2\tau$. The $L=2$ member of the quasi-$\gamma_1$ band 
is degenerate with the $L=4$ member of the gsb, the $L=3$, 4 members of the quasi-$\gamma_1$ band 
are degenerate to the $L=6$ member of the gsb, the $L=5$, 6 members of the quasi-$\gamma_1$ band 
are degenerate to the $L=8$ member of the gsb, and so on.  

\subsection{Axially symmetric prolate deformed nuclei}

In order to achieve exact separation of variables, we assume  a potential of the form \cite{Wilets,F1,F2,F3,ESDPRC}
\begin{equation}\label{eq:e7b}
v(\beta,\gamma)= u(\beta)+{f^2 \over \beta^2} w(\gamma),
\end{equation}
with $w(\gamma)$ having a deep minimum at $\gamma=0$.
Then the angular momentum term can be written as \cite{IacX5}
\begin{eqnarray}\label{Qdef}
\sum_{k=1,2,3} {Q_k^2 \over \sin^2\left( \gamma -{2\over 3} \pi k\right)} \nonumber \\
\approx {4\over 3} (Q_1^2+Q_2^2+Q_3^2)+Q_3^2\left( {1\over \sin^2\gamma}
-{4\over 3} \right).
\end{eqnarray}
One then seeks wave functions of the form \cite{IacX5}
\begin{equation}
\Psi(\beta,\gamma,\theta_i)= \phi_K^L(\beta,\gamma) 
{\cal D}^L_{M,K}(\theta_i), 
\end{equation}
where ${\cal D}(\theta_i)$ denote Wigner functions of the Euler angles, 
$L$ is the angular momentum quantum number, while $M$ and $K$ are the 
quantum numbers of the projections of angular momentum on the laboratory-fixed 
$z$-axis and the body-fixed $z'$-axis respectively. 
Then separation of variables leads to 
\begin{multline}\label{eq:e8}
\left[ 
-{1\over 2} {\sqrt{f}\over \beta^4} {\partial \over \partial \beta} 
\beta^4 f {\partial \over \partial \beta} \sqrt{f}
+{f^2\over 2\beta^2} \tilde \Lambda 
+{1\over 4} (1-\delta-\lambda) f\nabla^2 f \right. \\ \left.  
+{1\over 2} \left( {1\over 2}-\delta\right) \left( {1\over 2} -\lambda\right)
(\nabla f)^2 +u(\beta) \right] \xi_L(\beta) \\
= \epsilon \xi_L(\beta),  
\end{multline}
\begin{eqnarray}\label{eq:e9}
\left[ -{1\over \sin 3\gamma} {\partial \over \partial \gamma} \sin 3\gamma
{\partial \over \partial \gamma} + {K^2\over 4} \left({1\over \sin^2\gamma}-
{4\over 3}\right) \right.\nonumber \\
\left.  +2w(\gamma) \right] \eta_K (\gamma)
= \Lambda \eta_K(\gamma), 
\end{eqnarray}
where 
\begin{equation}
\tilde \Lambda = \Lambda+{L(L+1)\over 3},
\end{equation}
and  $\phi^L_K(\beta,\gamma)= \xi_L(\beta) \eta_K(\gamma).$
We remark that Eq. (\ref{eq:e8}) has the same form as Eq. (\ref{eq:e6}),
obtained in the case of $\gamma$-unstable nuclei, when $\tilde \Lambda$ in the  former is replaced 
by $\Lambda$ in the latter. However, the results are different as far as the physics described 
is concerned. The angular momentum dependence, contained in $\tilde \Lambda$ and $\Lambda$ respectively, is different.
Furthermore, the angular equation is different in each case, due to the different treatment of the $\gamma$ variable, 
the potential being confined to $\gamma \approx 0$ in the former case,  while being independent of $\gamma$ in the latter. 

Eq. (\ref{eq:e9}) has been solved for a harmonic oscillator potential 
\begin{equation}\label{wgamma}
w(\gamma)= {1\over 2} (3c)^2 \gamma^2 ,
\end{equation}
in the case of $\gamma \approx 0$ \cite{IacX5,ESDPRC},
resulting in 
\begin{equation}
\Lambda = \epsilon_\gamma -{K^2 \over 3}, \quad \epsilon_\gamma=
(6c) (n_\gamma+1), \quad n_\gamma=0,1,2,\dots,
\end{equation}
where $n_\gamma$ is the quantum number related to $\gamma$-oscillations. 
The allowed bands are characterized by 
\begin{eqnarray}
n_\gamma=0,\quad K=0; \qquad n_\gamma=1, \quad K=\pm 2;  \nonumber \\
n_\gamma=2, \quad K=0,\pm 4; \qquad \ldots
\end{eqnarray}
As a result 
\begin{equation}\label{Ltilde}
\tilde \Lambda = {L(L+1)-K^2\over 3}+\epsilon_\gamma = {L(L+1)-K^2\over 3}+ (6c) (n_\gamma+1).
\end{equation}

\subsection{Triaxial nuclei with $\gamma=\pi/6$} 

In this case we assume  again a potential of the form of Eq. (\ref{eq:e7b}), 
but  with $w(\gamma)$ having a deep minimum at $\gamma=\pi/6$.
In this case $K$, the angular momentum projection on the body-fixed 
$z'$-axis, is not a good quantum number any more, but $\alpha$, the angular 
momentum projection on the body-fixed $x'$-axis, is a good quantum number,
as found \cite{MtV} in the study of the triaxial rotator \cite{DF,DR}.  
Then the angular momentum term can be written as \cite{MtV,Z5}
\begin{equation}\label{Qtri}
\sum_{k=1,2,3} {Q_k^2 \over \sin^2\left( \gamma -{2\over 3} \pi k\right)}
\approx 4 (Q_1^2+Q_2^2+Q_3^2)-3 Q_1^2.
\end{equation}
One then seeks wave functions of the form \cite{Z5}
\begin{equation}
\Psi(\beta,\gamma,\theta_i)= \phi_\alpha^L(\beta,\gamma) 
{\cal D}^L_{M,\alpha}(\theta_i),
\end{equation}
where ${\cal D}(\theta_i)$ denote Wigner functions of the Euler angles, 
$L$ is the angular momentum quantum number, while $M$ and $\alpha$ are the 
quantum numbers of the projections of angular momentum on the laboratory-fixed 
$z$-axis and the body-fixed $x'$-axis respectively. 
Then separation of variables leads to 
\begin{multline}\label{eq:e8c}
\left[ 
-{1\over 2} {\sqrt{f}\over \beta^4} {\partial \over \partial \beta} 
\beta^4 f {\partial \over \partial \beta} \sqrt{f}  
+{f^2\over 2\beta^2} \bar \Lambda 
+{1\over 4} (1-\delta-\lambda) f\nabla^2 f \right. \\ \left.  
+{1\over 2} \left( {1\over 2}-\delta\right) \left( {1\over 2} -\lambda\right)
(\nabla f)^2 +u(\beta) \right] \xi_{L,\alpha}(\beta) \\
= \epsilon \xi_{L,\alpha}(\beta),  
\end{multline}
\begin{equation}\label{eq:e9c}
\left[ -{1\over \sin 3\gamma} {\partial \over \partial \gamma} \sin 3\gamma
{\partial \over \partial \gamma} +2w(\gamma) \right] \eta (\gamma)
= \Lambda' \eta(\gamma),
\end{equation}
with $\phi^L_\alpha (\beta,\gamma)= \xi_{L,\alpha}(\beta) \eta(\gamma)$,
and
\begin{equation}
\bar \Lambda = {4L(L+1)-3 \alpha^2 \over 4} + \Lambda'.
\end{equation}
Eq. (\ref{eq:e9c}) has been solved for a harmonic oscillator potential 
\begin{equation}
w(\gamma)= {1\over 4} c \left( \gamma -{\pi\over 6}\right)^2  ,
\end{equation}
in the case of $\gamma \approx \pi/6$ \cite{Z5},
resulting in 
\begin{equation}
\Lambda' = \epsilon_\gamma= \sqrt{2c} \left(n_\gamma+{1\over 2}\right), 
\end{equation}
where $n_\gamma$ is the quantum number related to $\gamma$-oscillations.
As a result
\begin{equation}\label{Lbar}
\bar \Lambda = {4L(L+1)-3 \alpha^2 \over 4} + \sqrt{2c} \left(n_\gamma+{1\over 2}\right).
\end{equation}

We remark that Eqs. (\ref{eq:e8}) and (\ref{eq:e8c}) have the same form, with $\tilde \Lambda$ 
in the former replaced by $\bar \Lambda$ in the latter. 

In the literature on triaxial nuclei it is customary, instead of the projection $\alpha$ of the angular momentum on the 
$x'$-axis, to introduce the wobbling quantum number 
\cite{BM,MtV} $n_w=L-\alpha$. Inserting $\alpha=L-n_w$ in 
Eq. (\ref{Lbar}) one obtains 
\begin{equation}\label{Lbar2}
\bar \Lambda = {L(L+4)+3 n_w(2L-n_w) \over 4} + \sqrt{2c} \left(n_\gamma+{1\over 2}\right).
\end{equation}

\subsection{Common form of the radial equation}\label{3cases} 

We remark that Eqs. (\ref{eq:e6}), (\ref{eq:e8}), and (\ref{eq:e8c}) have the same form, 
the only difference being that $\Lambda$ in the first equation is replaced by  $\tilde \Lambda$ 
in the second, and  by $\bar \Lambda$ in the third one. In what follows we are going to use the symbol $\Lambda$,
understanding that 

i) for $\gamma$-unstable nuclei it is given by $\Lambda= \tau(\tau+3)$,

ii) for axially symmetric prolate deformed nuclei it should be replaced by $\tilde \Lambda$, given in Eq. (\ref{Ltilde}), and 

iii) for triaxial nuclei it should be replaced by $\bar \Lambda$, given in Eq. (\ref{Lbar2}).

Eq. (\ref{eq:e6}) can be simplified by performing the derivations
\begin{eqnarray}\label{eq:e10}
& {1\over 2 } f^2 \xi''+ \left( f f'+{2 f^2\over \beta}\right) \xi' 
+ \left( {(f')^2\over 8} + {f f''\over 4} +{f f'\over \beta}\right) \xi \nonumber \\
&  -{f^2 \over 2\beta^2}  \Lambda \xi +\epsilon \xi -v_{eff}\xi =0, 
\end{eqnarray}
with
\begin{eqnarray}
v_{eff}= u + {1\over 4} (1-\delta-\lambda)  f \left( {4 f'\over \beta} 
+f''\right) \nonumber \\ 
+ {1\over 2} \left( {1\over 2}-\delta\right) 
\left( {1\over 2}-\lambda\right) (f')^2. 
\end{eqnarray}
The difference in the numerical coefficient of $f'$ observed in comparison 
to Eq. (2.27) of Ref. \cite{QT4267} is due to the different dimensionality 
of the space used in each case. 

Setting 
\begin{equation}
\xi(\beta)= {R(\beta)\over \beta^2}, 
\end{equation}
Eq. (\ref{eq:e10}) is put into the form 
\begin{equation}\label{eq:e14}
H R= - \left(\sqrt{f}{d\over d\beta}\sqrt{f}\right)^2 R + 2 u_{eff} R 
=2 \epsilon R, 
\end{equation}
where 
\begin{equation}\label{eq:e15}
u_{eff}= v_{eff} + {f^2+\beta f f'\over \beta^2}+ {f^2 \over 2 \beta^2}  \Lambda.   
\end{equation}

\section{The Davidson potential} 

Up to now no assumption about the specific form of the potential 
$u(\beta)$ and the deformation function $f(\beta)$ has been made.
We are now going to consider the special case of the Davidson potential \cite{Dav}
\begin{equation} \label{eq:e16}
u(\beta)=\beta^2 + {\beta_0^4\over \beta^2},
\end{equation}
where the parameter $\beta_0$ indicates the position of the minimum 
of the potential. The special case of $\beta_0=0$ corresponds to the 
simple harmonic oscillator. 

Based on the results for the 3-dimensional harmonic oscillator reported 
in Ref. \cite{Q2929}, we are also going to consider for the deformation 
function the special form 
\begin{equation}\label{eq:e17}
f(\beta)=1+a \beta^2, \qquad a \ll 1.
\end{equation}
This choice is made in order to lead to an exact solution. Its physical 
implications will be discussed in Section 11.

Using these forms for the potential and the deformation function 
in Eq. (\ref{eq:e15}) one obtains 
\begin{equation}\label{Ueffect}
 2 u_{\rm eff} = k_1 \beta^2 + k_0 + \frac{k_{-1}}{\beta^2},
\end{equation} 
where 
\begin{equation}\label{eq:ueff}
\begin{split}
  & k_1 = 2 + a^2 [5(1 - \delta - \lambda) + (1 - 2\delta) (1 - 2\lambda) + 6 + \Lambda ], \\
  & k_0 = a [5 (1 - \delta - \lambda) + 8 + 2\Lambda ], \\
  & k_{-1} = 2 + \Lambda + 2 \beta_0^4.
\end{split}
\end{equation}

\section{Deformed shape invariance}

Our task now is to find the eigenvalues and eigenfunctions of the Hamiltonian of Eq. (\ref{eq:e14}). 
This can be achieved by imposing shape invariance \cite{Q2929}, which is an integrability condition guaranteeing 
that exact solutions of the Hamiltonian of Eq. (\ref{eq:e14}) can be found. The use of shape invariance in the 
framework of SUSYQM is equivalent \cite{SUSYQM} to the well known factorization method of the Schr\"odinger equation, 
introduced 60 years ago by Infeld and Hull \cite{Infeld}. In other words, we are now going to use 
a mathematical technique allowing us to find the solutions of Eq. (\ref{eq:e14}).

In its simplest form in an one-dimensional space, shape invariance can be described as follows \cite{Balantekin}. 
Two potentials $V_1$ and $V_2$, which are supersymmetric partners, are in general different functions of $x$. 
They are called shape invariant if they satisfy the condition 
\begin{equation}\label{Bal1}
V_2(x;a_1) = V_1(x;a_2)+R(a_1),
\end{equation}
where $a_1$, $a_2$ are sets of parameters independent of $x$, with $a_2$ being a function of $a_1$, and the remainder $R(a_1)$
is also independent of $x$. In other words, the two potentials have the same functional dependence on $x$, the difference 
being in the values of the parameters appearing in each of them, and in their relative displacement by the remainder $R(a_1)$. 
Furthermore, it is known that the shape invariance condition of Eq. (\ref{Bal1}) can be written in the operator form
\begin{equation}
A(a_1) A^\dagger(a_1)= A^\dagger(a_2) A(a_2) +R(a_1), 
\end{equation} 
where $A$ and $A^\dagger$ are the operators corresponding to the supersymmetric partners $H_1=A^\dagger A$ and 
$H_2= A A^\dagger$. Solving the Schr\"odinger equation for $H_1$ by this method, one obtains as a ``bonus'' the solution 
of $H_2$ as well. 

In the present case, the concept of shape invariance has to be generalized, as described in detail in Ref. \cite{Q2929},
since the mass depends on the deformation, resulting in a deformed shape invariance condition. Instead of two Hamiltonians,
one has a series of many Hamiltonians. We are interested in solving the Schr\"odinger equation for the first of them, 
which will be Eq. (\ref{eq:e14}).

$H$ in Eq. (\ref{eq:e14}) may be considered as the first member $H_0 = H$ of a hierarchy of Hamiltonians
\begin{equation}
  H_i = A_i^+ A_i^- + \sum_{j=0}^i \varepsilon_j, \qquad i=0, 1, 2, \ldots,
\end{equation}
where the first-order operators \cite{Q2929}
\begin{equation}
  A_i^{\pm} = A^{\pm}(\mu_i, \nu_i) = \mp \sqrt{f} \frac{d}{d\beta} \sqrt{f} + W(\mu_i, 
  \nu_i; \beta) 
\end{equation}
satisfy a deformed shape invariance condition
\begin{equation}
  A_i^- A_i^+ = A_{i+1}^+ A_{i+1}^- + \varepsilon_{i+1}, \qquad i=0, 1, 2, \ldots,
\end{equation}
with $\varepsilon_i$, $i=0$, 1, 2,~\ldots, denoting some constants. 
(Note that the parameters $\lambda$ and $\mu$ of \cite{Q2929} have been changed into $\mu$ and $\nu$, respectively.)

In other words, the superpotential $W(\mu, \nu; \beta)$ fulfils the two conditions
\begin{equation}
  W^2(\mu, \nu; \beta) - f(\beta) W'(\mu, \nu; \beta) + \varepsilon_0 = 2 u_{\rm eff}(\beta) 
  \label{eq:SI-1}
\end{equation}
and
\begin{multline}
  W^2(\mu_i, \nu_i; \beta) + f(\beta) W'(\mu_i, \nu_i; \beta) \\ = W^2(\mu_{i+1}, 
        \nu_{i+1}; \beta) - f(\beta) W'(\mu_{i+1}, \nu_{i+1}; \beta) + \varepsilon_{i+1}, \\
  i=0, 1, 2, \ldots,  \label{eq:SI-2}
\end{multline}
where $\mu_0 = \mu$, $\nu_0 = \nu$, and a prime denotes derivative with respect to $\beta$.

In the case of the effective potential given in Eq.~(\ref{Ueffect}), $W(\mu, \nu; \beta)$ is a class 2 superpotential
\begin{align}
  & W(\mu, \nu; \beta) = \mu \phi(\beta) + \frac{\nu}{\phi(\beta)}, \label{eq:W}\\
  & \phi(\beta) = \frac{1}{\beta},  \label{eq:phi}
\end{align}
which means that Eqs.~(3.9) and (3.10) of \cite{Q2929} read
\begin{equation}
\begin{split}
  & \phi'(\beta) = - \frac{1}{\beta^2} = \frac{A}{\beta^2} + B, \\
  & a \beta^2 = \frac{(A'/\beta^2) + B'}{(-1/\beta^2)},
\end{split}
\end{equation}
with $A = -1$, $B = 0$, $A' = 0$, and $B' = - a$.

Inserting Eqs.~(\ref{eq:W}) and (\ref{eq:phi}) in (\ref{eq:SI-1}), we obtain
\begin{equation}
\begin{split}
&  \left(\frac{\mu}{\beta} + \nu \beta\right)^2 - (1 + a \beta^2) \left(- \frac{\mu}{\beta^2} 
  + \nu\right) + \varepsilon_0 \\
&  = k_1 \beta^2 + k_0 + \frac{k_{-1}}{\beta^2}, 
\end{split}
\end{equation}
which is equivalent to the three equations
\begin{equation}
  \mu (\mu+1) = k_{-1}, \quad \nu (\nu - a) = k_1, \quad 2 \mu \nu + \mu a -
  \nu + \varepsilon_0 = k_0.
\end{equation}
Their solutions read
\begin{eqnarray}
  \mu = \frac{1}{2} (- 1 \pm \Delta_1), \qquad \nu = \frac{a}{2} (1 \pm \Delta_2), \nonumber \\
         \varepsilon_0 = k_0 - 2\mu\nu - \mu a + \nu, \nonumber \\
   \Delta_1 \equiv \sqrt{1 + 4 k_{-1}}, \qquad \Delta_2 \equiv \sqrt{1 + 4 \frac{k_1}{a^2}},  
   \label{eq:Delta}
\end{eqnarray}
provided $1 + 4 k_1/a^2 \ge 0$ (note that $1 + 4 k_{-1}$ is always positive). As we shall show in Sec. \ref{gswf}, 
the conditions ensuring that the ground-state wavefunction is physically acceptable select 
the lower sign for $\mu$ and the upper one for $\nu$:
\begin{equation}
  \mu = - \frac{1}{2} (1 +\Delta_1), \qquad \nu = \frac{a}{2} (1 + \Delta_2).  
  \label{eq:lambda-mu}
\end{equation}

Inserting next Eqs.~(\ref{eq:W}) and (\ref{eq:phi}) in Eq.~(\ref{eq:SI-2}), we get
\begin{equation} \begin{split}
 & \left(\frac{\mu_i}{\beta} + \nu_i \beta\right)^2 + (1 + a \beta^2) \left(- \frac{\mu_i}
  {\beta^2} + \nu_i\right)  \\
 &  = \left(\frac{\mu_{i+1}}{\beta} + \nu_{i+1} \beta\right)^2 - (1 + a 
  \beta^2) \left(- \frac{\mu_{i+1}}{\beta^2} + \nu_{i+1}\right)  \\
 &  + \varepsilon_{i+1},
\end{split}
\end{equation}
leading to the three conditions
\begin{equation}
\begin{split}
  & \mu_i (\mu_i - 1) = \mu_{i+1} (\mu_{i+1} + 1), \\ 
  & \nu_i (\nu_i + a) = \nu_{i+1} (\nu_{i+1} - a), \\
  & 2 \mu_i \nu_i - \mu_i a + \nu_i = 2 \mu_{i+1} \nu_{i+1} + \mu_{i+1} a -
       \nu_{i+1} + \varepsilon_{i+1}.  
\end{split}
\end{equation}
Their solutions are
\begin{equation}
  \mu_{i+1} = \mu_i - 1, \qquad \nu_{i+1} = \nu_i + a,  \label{eq:sol-SI-2}
\end{equation}
and
\begin{equation}
  \varepsilon_{i+1} = 2 (\mu_i \nu_i - \mu_{i+1} \nu_{i+1}) - (\mu_i + \mu_{i+1}) a
  + \nu_i  + \nu_{i+1}.
\end{equation}
Note that there are other solutions for $\mu_{i+1}$ and $\nu_{i+1}$, namely $\mu_{i+1} = - \mu_i$ and $\nu_{i+1} = - \nu_i$, but the alternating signs would not be compatible with physically acceptable excited-state wavefunctions. Finally, the iteration of (\ref{eq:sol-SI-2}) leads to
\begin{equation}
  \mu_i = \mu - i, \qquad \nu_i = \nu + ia.  \label{eq:lambda-mu-i}
\end{equation}

\section{Energy spectrum}

The energy spectrum of Eq.~(\ref{eq:e14}) is therefore given by
\begin{equation}
\begin{split}
 & \epsilon_n  = \tfrac{1}{2} \sum_{i=0}^n \varepsilon_i \\ 
  & = \tfrac{1}{2}\left[k_0 - 2 \mu_n \nu_n - a \left(2 \sum_{i=0}^{n-1} \mu_i +
        \mu_n\right) + 2 \sum_{i=0}^{n-1} \nu_i + \nu_n\right] \\
  & = \tfrac{1}{2} [k_0 - 2 \mu \nu - a \mu + \nu - 4 (a\mu - \nu) n + 4a n^2].
\end{split}
\end{equation}
On taking (\ref{eq:lambda-mu}) into account, this can be rewritten as
\begin{equation}
\begin{split}\label{eq:energy}
 & \epsilon_n = \tfrac{1}{2} [k_0 + \tfrac{1}{2} a (3 + 2\Delta_1 + 2\Delta_2 + \Delta_1 \Delta_2) \\
 & + 2a (2 + \Delta_1 + \Delta_2) n + 4a n^2], \qquad n=0, 1, 2, \ldots.  
\end{split}
\end{equation}
Equation (\ref{eq:energy}) only provides a formal solution to the bound-state energy spectrum. 
The range of $n$ values is actually determined by the existence of corresponding physically acceptable wavefunctions.
The relevant conditions will be considered in the next section. 

We  quote here the final results for the spectra, which will be used for comparison to experiment. 
One has 
\begin{equation}\label{Egsb}
\begin{split}
& \epsilon_0 = {19\over 4} a +{5\over 2}(1-\delta-\lambda) a
+ {1\over 2} \sqrt{a^2+4 k_1} \\ 
& + {a\over 2} \sqrt{1+4 k_{-1}}  + {1\over 4} \sqrt{(a^2+4 k_1)(1 + 4 k_{-1})}
+ a \Lambda,
\end{split}
\end{equation}
\begin{equation}\label{Eb1}
\epsilon_1= \epsilon_0 + 4 a + \sqrt{a^2+4 k_1} + a \sqrt{1+4 k_{-1}},
\end{equation}
\begin{equation}\label{Eb2}
\epsilon_2= \epsilon_0 + 12 a + 2\sqrt{a^2+4 k_1} + 2a \sqrt{1+4 k_{-1}},
\end{equation}
where $k_1$, $k_{-1}$ are given by Eq. (\ref{eq:ueff}), in which $\Lambda$
has the form explained in subsec. \ref{3cases}.  

The ground state band is obtained from Eq. (\ref{Egsb}), while the quasi-$\beta_1$ band 
is obtained from Eq. (\ref{Eb1}), and the quasi-$\beta_2$ band is obtained from Eq. (\ref{Eb2}).

In the special case of $a=0$ (no dependence of the mass on the deformation) one easily obtains
\begin{equation}
\epsilon_1= \epsilon_0 + 2\sqrt{2}, \qquad \epsilon_2= \epsilon_0 + 4\sqrt{2},
\end{equation}
i.e. the $\beta$-bandheads become equidistant. 

\section{Wave functions} 

To be physically acceptable, the bound-state wavefunctions should satisfy two conditions \cite{Q2929}:\par
\noindent (i) As in conventional (constant-mass) quantum mechanics, they should be square integrable on
the interval of definition of $u_{\rm eff}$, i.e.,
\begin{equation}
  \int_{0}^{\infty} d\beta\,  |R_n(\beta)|^2 < \infty.  \label{eq:wf-C1}
\end{equation}
\noindent (ii) Furthermore, they should ensure the Hermiticity of $H$. For such a purpose, it is enough to impose that the operator $\sqrt{f} (d/d\beta) \sqrt{f}$ be Hermitian, which amounts to the restriction
\begin{equation}
  |R_n(\beta)|^2 f(\beta) \to 0 \qquad {\rm for\ } \beta \to 0 {\rm \ and\ } \beta \to \infty,
\end{equation}
or, equivalently,
\begin{eqnarray}
  |R_n(\beta)|^2 \to 0 \quad {\rm for\ } \beta \to 0 \qquad {\rm and} \nonumber \\
  |R_n(\beta)|^2 \beta^2  \to 0 \quad {\rm for\ } \beta \to \infty.  \label{eq:wf-C2}
\end{eqnarray}
As condition (\ref{eq:wf-C2}) is more stringent than condition (\ref{eq:wf-C1}), we should only be concerned with the former.

\subsection{Ground-state wavefunction}\label{gswf}

The ground-state wavefunction, which is annihilated by $A^-$, is given by Eq.~(2.29) of \cite{Q2929} as
\begin{eqnarray}
  R_0(\beta) = R_0(\mu, \nu; \beta) \nonumber \\
  = \frac{N_0}{\sqrt{f(\beta)}} \exp \left(- \int^{\beta}
  \frac{W(\mu, \nu; \tbeta)}{f(\tbeta)} d\tbeta\right),
\end{eqnarray}
where $N_0$ is some normalization coefficient. Here
\begin{eqnarray}
  \int^{\beta} \frac{W(\mu, \nu; \tbeta)}{f(\tbeta)} d\tbeta = \int^{\beta} \left(\frac{\mu}
  {\tbeta} + \frac{(\nu - \mu a) \tbeta}{1 + a \tbeta^2}\right) d\tbeta \nonumber \\
  = \mu \ln \beta + \frac{1}{2a} (\nu - \mu a) \ln (1 + a \beta^2).
\end{eqnarray}
Hence
\begin{equation}
  R_0(\beta) = N_0 \beta^{-\mu} f^{- (\nu - \mu a + a)/(2a)}.  \label{eq:R_0}
\end{equation}

{}For $\beta \to 0$, the function $|R_0(\beta)|^2$ behaves as $\beta^{-2\mu}$. Condition (\ref{eq:wf-C2}) imposes that $- 2\mu > 0$ or $\mu < 0$. Since $k_{-1}$, defined in Eq.~(\ref{eq:ueff}), is greater than 2, it follows that $\Delta_1$, defined in (\ref{eq:Delta}), is greater than 3, so that the upper sign choice for $\mu$ in (\ref{eq:Delta}) would lead to $\mu > 1$. As this is not acceptable, we have to take the lower sign for which $\mu < -2$.\par

{}For $\beta \to \infty$, $|R_0(\beta)|^2 \beta^2$ behaves as $\beta^{- 2\nu/a}$. Condition (\ref{eq:wf-C2}) therefore imposes that $\nu > 0$. This restriction is surely satisfied by the upper sign choice for $\nu$ in (\ref{eq:Delta}). 
For the lower one, it is not fulfilled if we restrict ourselves to small enough values of $a$ because then $k_1$ in (\ref{eq:ueff}) will be positive and $\Delta_2$ in (\ref{eq:Delta}) will be greater than 1. For sufficiently large values of $a$, however, both sign choices might be acceptable. Since among two acceptable wavefunctions, it is customary in quantum mechanics to choose the most regular one (see, e.g., \cite{Znojil} and references quoted therein), we assume the upper sign for $\nu$, thus getting Eq.~(\ref{eq:lambda-mu}).

\subsection{Excited-state wavefunctions}

According to Eqs.~(2.30), (3.20) and (3.21) of \cite{Q2929}, the excited-state wavefunctions are given by
\begin{equation}
\begin{split}
&  R_n(\beta) = R_n(\mu, \nu; \beta) \propto \beta^{-n} R_0(\mu_n, \nu_n; \beta) 
  P_n(\mu, \nu; y), \\
&   \qquad y = \beta^2,  \label{eq:R_n}
   \end{split}
\end{equation}
where $P_n(\mu, \nu; y)$ is an $n$th-degree polynomial in $y$, satisfying the equation
\begin{equation}
\begin{split}
&  P_{n+1}(\mu, \nu; y) = - 2y (1 + ay) \frac{d}{dy} P_n(\mu_1, \nu_1; y) \\  
&  + [\mu_{n+1}  + \mu + n + (\nu_{n+1} + \nu + na) y] P_n(\mu_1, \nu_1; y),  \label{eq:P-eq}
\end{split}
\end{equation}
with the starting value $P_0(\mu, \nu; y) = 1$.\par

{}From Eqs.~(\ref{eq:lambda-mu-i}) and (\ref{eq:R_0}), it follows that 
\begin{equation}
\begin{split}
&  R_0(\mu_n, \nu_n; \beta) \propto \beta^{-\mu_{n}} f^{-(\nu_n - \mu_n a + a)/(2a)} \\ 
&  \propto R_0(\mu, \nu; \beta) \beta^n f^{-n},
\end{split}
\end{equation}
so that Eq.~(\ref{eq:R_n}) becomes
\begin{equation}
  R_n(\beta) \propto R_0(\beta) f^{-n} P_n(\mu, \nu; y).
\end{equation}
It is then clear that $R_n(\beta)$ satisfies condition (\ref{eq:wf-C2}) for any $n=1$, 2,~\ldots, since $R_0(\beta)$ does.\par

It now remains to solve Eq.~(\ref{eq:P-eq}). For such a purpose, let us make the changes of variable and of function
\begin{equation}
\begin{split}
&  t = 1 - \frac{2}{f} = \frac{- 1 + ay}{1 + ay}, \\
&  P_n(\mu, \nu; y) = C_n f^n  Q_n(\mu, \nu; t),  \label{eq:change}
\end{split}
\end{equation}
where $C_n$ is some constant. From definition (\ref{eq:change}), it follows that $Q_n(\mu, \nu; t)$ an $n$th-degree polynomial in $t$. We successively get
\begin{equation}
  y = \frac{1+t}{a(1-t)}, \quad 1 + ay = \frac{2}{1-t}, \quad \frac{d}{dy} = \frac{a}{2} (1-t)^2
  \frac{d}{dt}.
\end{equation}
It is then straightforward to show that Eq.~(\ref{eq:P-eq}) becomes
\begin{equation}
\begin{split}
 & \frac{C_{n+1}}{C_n} Q_{n+1}(\mu, \nu; t) 
 = \left\{ - (1 - t^2) \frac{d}{dt} \right. \\ 
 & \left. +   \left[\mu + \frac{\nu}{a} + \left(\frac{\nu}{a} - \mu + 1\right) t\right]\right\}
  Q_n(\mu - 1, \nu + a; t).
\end{split}
\end{equation}

On taking into account that the Jacobi polynomials satisfy the backward shift operator relation (see Eq.~(1.8.7) of \cite{Koekoek})
\begin{equation}
\begin{split}
&  2(n+1) P_{n+1}^{(\alpha, \beta)} (x) = \left\{- (1 - x^2) \frac{d}{dx} \right.\\
& \left. + [\alpha - \beta + (\alpha +
  \beta + 2) x]\right\} P_n^{(\alpha+1, \beta+1)}(x),
\end{split}
\end{equation}
we see that $Q_n(\mu, \nu; t)$ is actually some Jacobi polynomial
\begin{equation}
  Q_n(\mu, \nu; t) = P_n^{\left(\frac{\nu}{a} - \frac{1}{2}, - \mu - \frac{1}{2}\right)}(t)
  = P_n^{\left(\frac{\Delta_2}{2}, \frac{\Delta_1}{2}\right)}(t),
\end{equation}
provided we choose
\begin{equation}
  \frac{C_{n+1}}{C_n} = 2 (n+1), \qquad C_0 = 1,
\end{equation}
or, in other words, $C_n = 2^n n!$.\par

We therefore conclude that the wavefunctions are given by
\begin{equation}
\begin{split}
&  R_n(\beta) = \frac{N_n}{N_0} R_0(\beta) P_n^{\left(\frac{\nu}{a} - \frac{1}{2}, - \mu - 
  \frac{1}{2}\right)}(t) \\
&  = N_n \beta^{- \mu} f^{- (\nu - \mu a + a)/(2a)} P_n^{\left(
  \frac{\nu}{a} - \frac{1}{2}, - \mu - \frac{1}{2}\right)}(t),
\end{split}
\end{equation}
or
\begin{equation}\label{R_n}
\begin{split}
 & R_n(\beta) = N_n \beta^{(1 + \Delta_1)/2} f^{- 1 - (\Delta_1 + \Delta_2)/4} P_n^{(\Delta_2/2,
  \Delta_1/2)}(t), \\
  & t = \frac{- 1 + a\beta^2}{1 + a \beta^2},
\end{split}
\end{equation}
where $N_n$ is some normalization coefficient.

The Jacobi polynomials appearing in the wave functions of the ground state band ($n=0$), 
the quasi-$\beta_1$ band ($n=1$), and the  quasi-$\beta_2$ band ($n=2$),
needed for the calculation of the relevant $B(E2)$ transitions, read
\begin{equation}
 P_0^{(\alpha,\beta)}(x)=1, 
 \end{equation}
 \begin{equation}
 P_1^{(\alpha,\beta)}(x)={1\over 2} [ 2(\alpha+1)+(\alpha+\beta+2)(x-1)],
 \end{equation}
 \begin{equation}
 \begin{split}
& P_2^{(\alpha,\beta)}(x)={1\over 8} [ 4(\alpha+1)(\alpha+2) \\
& +4 (\alpha+\beta+3)(\alpha+2)(x-1) \\
& + (\alpha+\beta+3)(\alpha+\beta+4)(x-1)^2]. 
\end{split}
\end{equation}

\section{Normalization coefficient}

To calculate $N_n$, let us first express the whole wavefunction $R_n$ in terms of $t$:
\begin{equation}
\begin{split}
  R_n &= N_n y^{(1 + \Delta_1)/4} (1 + ay)^{- 1 - (\Delta_1 + \Delta_2)/4} P_n^{(\Delta_2/2,
       \Delta_1/2)}(t)  \\
  &= N_n 2^{- 1 - (\Delta_1 + \Delta_2)/4} a^{- (1 + \Delta_1)/4} (1+t)^{(1 + \Delta_1)/4} \\
   &    (1-t)^{(3 + \Delta_2)/4} P_n^{(\Delta_2/2, \Delta_1/2)}(t).  
\end{split}
\end{equation}
Now, on taking into account that 
\begin{equation}
  d\beta = \frac{dy}{2\sqrt{y}} = \frac{dt}{\sqrt{a} (1-t)^{3/2} (1+t)^{1/2}},
\end{equation}
we obtain
\begin{equation}
\begin{split}
&  \int_0^{\infty} |R_n|^2 d\beta = |N_n|^2 2^{- 2 - (\Delta_1 + \Delta_2)/2} a^{- 1 - \Delta_1/2} \\ 
&  \int_{-1}^{+1} (1-t)^{\Delta_2/2} (1+t)^{\Delta_1/2} \left[P_n^{(\Delta_2/2, \Delta_1/2)}(t) \right]^2 dt
  \end{split}
\end{equation}
in terms of the normalization integral of Jacobi polynomials \cite{AbrSte}.

Hence the normalization condition reads
\begin{equation}
\begin{split}
&  |N_n|^2 2^{- 2 - (\Delta_1 + \Delta_2)/2} a^{- 1 - \Delta_1/2} \\
&   \frac{2^{(\Delta_1 + \Delta_2)/2
  +1} \Gamma\left(n + \frac{\Delta_1}{2} + 1\right) \Gamma\left(n + \frac{\Delta_2}{2} + 1\right)}
  {\left(2n + \frac{\Delta_1 + \Delta_2}{2} + 1\right) n!\, \Gamma\left(n + \frac{\Delta_1 + 
  \Delta_2}{2} + 1\right)} = 1, 
\end{split}
\end{equation}
and leads to
\begin{equation}\label{normaliz}
\begin{split}
&  N_n = 
  \left(2 a^{\Delta_1/2 + 1} \left(2n + \frac{\Delta_1 + \Delta_2}{2} + 1\right) n!\right)^{1/2} \\ 
& \left(\frac{    \Gamma\left(n + \frac{\Delta_1 + \Delta_2}{2} + 1\right)}{\Gamma\left(n + \frac{\Delta_1}{2} + 1
  \right) \Gamma\left(n + \frac{\Delta_2}{2} + 1\right)}\right)^{1/2}. 
\end{split}
\end{equation}

A way of avoiding numerical problems when having to handle $\Gamma(x)$ functions with large $x$
is given in Appendix 1.

\section{$B(E2)$ transition rates} 

B(E2) transition rates
\begin{equation}
B(E2; \varrho L \to \varrho' L')= {5\over 16 \pi} { |\langle \varrho' L' || T^{(E2)}
|| \varrho L  \rangle|^2  \over 2L+1},
\end{equation}
where $\varrho$ stands for quantum numbers other than the angular momentum $L$, 
can be calculated using the quadrupole operator $T^{(E2)}$
and the Wigner-Eckart theorem in the form
\begin{equation}
\begin{split}
& \langle \varrho' L' M' | T^{(E2)}_{\mu} | \varrho L M  \rangle \\
& = {1\over \sqrt{2L'+1}} \langle L 2 L' | M \mu M'\rangle \langle \varrho' L'  ||
T^{(E2)} || \varrho L \rangle .
\end{split}
\end{equation}

\subsection{$B(E2)$s for $\gamma$-unstable nuclei}

The calculation is carried out exactly as in Ref. \cite{BonE5},
using the quadrupole operator 
\cite{IacE5}
\begin{equation}\label{eq:TE2}
\begin{split}
& T^{(E2)} = A \beta \left[ {\cal D}^{(2)}_{\mu,0}(\theta_i) \cos\gamma \right. \\
& \left. +{1\over \sqrt{2}} \left( {\cal D}^{(2)}_{\mu,2}(\theta_i)+ {\cal
D}^{(2)}_{\mu,-2}(\theta_i)\right) \sin\gamma \right],
\end{split}
\end{equation}
where $A$ is a scale factor. 

The results of Ref. \cite{BonE5} need not be repeated here. The only difference is that in the radial 
integral (see Eq. (21) of Ref. \cite{BonE5})
the wave functions $R_{n,\tau}(\beta)$ appear
\begin{equation}
\begin{split}
& I_{n',\tau+1; n,\tau} = \int_0^\infty \beta \xi_{n',\tau+1}(\beta) \xi_{n,\tau}(\beta) \beta^4 d\beta \\
& =\int_0^\infty \beta R_{n',\tau+1}(\beta) R_{n,\tau}(\beta) d\beta.
\end{split}
\end{equation}
The $\tau$ dependence of the wave functions $R_n(\beta)$ of Eq. (\ref{R_n}) is contained in 
$\Delta_1$, $\Delta_2$, known from Eq. (\ref{eq:Delta}) to contain $k_1$, $k_{-1}$, 
which in turn are known from Eq. (\ref{eq:ueff}) to contain $\Lambda=\tau(\tau+3)$.  

\subsection{$B(E2)$s for axially symmetric prolate deformed nuclei}

The quadrupole operator is again given by Eq. (\ref{eq:TE2}). 
The calculation is carried out exactly as in Ref. \cite{ESDPRC}, 
the results of which need not be repeated here. The only difference is that in the radial 
integral (see Eq. (B5) of Ref. \cite{ESDPRC})
the wave functions $R_{n,L}(\beta)$ appear
\begin{equation}\begin{split}
& B_{n,L,n',L'} = \int_0^\infty \beta \xi_{n,L}(\beta) \xi_{n',L'}(\beta) \beta^4 d\beta \\
& =\int_0^\infty \beta R_{n,L}(\beta) R_{n',L'}(\beta) d\beta.
\end{split}
\end{equation}
The $L$ dependence of the wave functions $R_n(\beta)$ of Eq. (\ref{R_n}) is contained in 
$\Delta_1$, $\Delta_2$, known from Eq. (\ref{eq:Delta}) to contain $k_1$, $k_{-1}$, 
which in turn are known from Eq. (\ref{eq:ueff}) to contain $\tilde \Lambda$ of Eq. (\ref{Ltilde}).  

\subsection{$B(E2)$s for triaxial nuclei with $\gamma=\pi/6$}

The calculation is carried out exactly as in Ref. \cite{Yig}, using the quadrupole operator
\begin{equation}\label{eq:e31}
\begin{split}
& T^{(E2)}_\mu = A \beta \left[ {\cal D}^{(2)}_{\mu,0}(\theta_i)\cos\left(\gamma
-{2\pi\over 3}\right) \right. \\
& \left.  +{1\over \sqrt{2}}
({\cal D}^{(2)}_{\mu,2}(\theta_i)+{\cal D}^{(2)}_{\mu,-2}(\theta_i) )
\sin\left(\gamma -{2\pi\over 3} \right) \right],
\end{split}
\end{equation}
where $A$ is a scale factor, while the quantity $\gamma -2\pi/3$ in the trigonometric functions
is obtained from $\gamma-2\pi k/3$ for $k=1$, since in the present case
the projection $\alpha$ along the body-fixed $\hat x'$-axis is used.

The results of Ref. \cite{Yig}  need not be repeated here. The only difference is that in the radial 
integral (see Eq. (14) of Ref. \cite{Yig}) the wave functions $R_{n,\alpha,L}(\beta)$ appear
\begin{equation}
\begin{split}
& I_\beta(n,L,\alpha;n',L',\alpha') = \int_0^\infty \beta \xi_{n,\alpha,L}(\beta)
 \xi_{n',L',\alpha'}(\beta) \beta^4 d\beta \\
 & =\int_0^\infty \beta R_{n,\alpha,L}(\beta) R_{n',\alpha',L'}(\beta) d\beta.
\end{split}
\end{equation}
The $\alpha,L$ dependence of the wave functions $R_n(\beta)$ of Eq. (\ref{R_n}) is contained in 
$\Delta_1$, $\Delta_2$, known from Eq. (\ref{eq:Delta}) to contain $k_1$, $k_{-1}$, 
which in turn are known from Eq. (\ref{eq:ueff}) to contain $\bar \Lambda$ of Eq. (\ref{Lbar}).  


\begin{figure}[ht]
\centering
\includegraphics[height=6cm ]{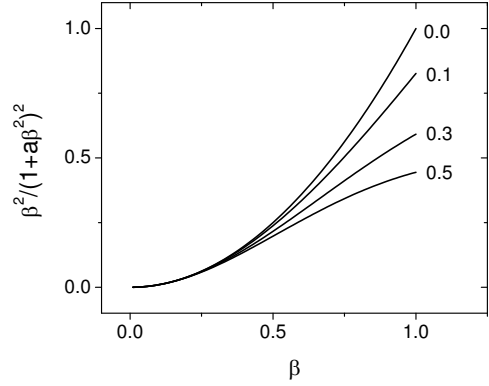}
\caption{The function $\beta^2 / f^2(\beta) = \beta^2 / (1+a \beta^2)^2$, to which 
moments of inertia are proportional as seen from Eq. (\ref{eq:mBohr}), plotted 
as a function of the nuclear deformation $\beta$ for different values of the parameter $a$.
See Section XI for further discussion.}
\end{figure}


\begin{figure}[ht]
\centering
\includegraphics[height=7cm ]{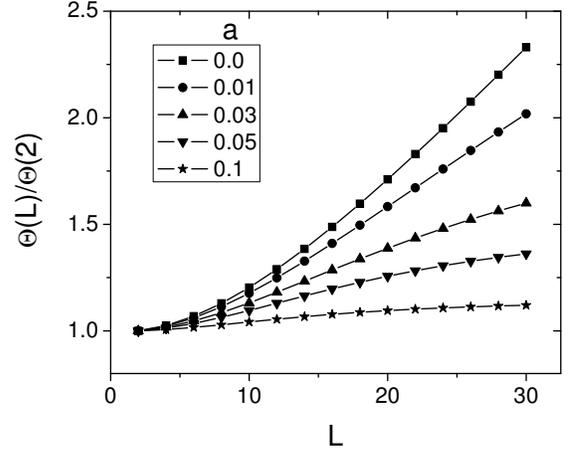}
\caption{Moments of inertia $\Theta(L)$ 
for the ground state band, given by Eq. (\ref{eq:Theta}) and normalized to  
$\Theta(2)$, are shown for the specific values of $\beta_0=2$ and $c=5$, and varying parameter $a$. 
See Section XI for further discussion.}
\end{figure}

\section{Numerical results} 

From Eq. (\ref{eq:mBohr}) it is clear that in the present case the moments of inertia 
are not proportional to $\beta^2 \sin^2\left( \gamma -2 \pi k/3\right)$ but to 
$(\beta^2/f^2(\beta)) \sin^2\left( \gamma -2 \pi k/3\right)$. The function $\beta^2/ 
f^2(\beta)$ is shown in Fig. 1 for different values of the parameter $a$. It is clear 
that the increase of the moment of inertia is slowed down by the function $f(\beta)$,
as it is expected as nuclear deformation sets in \cite{Ring}.  

The effect of the deformation-dependent mass on the moments of inertia can be seen in Fig.~2, where the moments of inertia \cite{Ring} 
for the ground state band 
\begin{equation}\label{eq:Theta}
\Theta(L)= {2L-1 \over E(L)-E(L-2)},
\end{equation}
normalized to $\Theta(2)$, are shown in the case of axially symmetric prolate deformed nuclei,
for the specific values of $\beta_0=2$ and $c=5$, and varying parameter $a$. 
It is clear that the rapid increase of the moments of inertia with $L$, seen for $a=0$, is gradually moderated by increasing $a$.

\subsection{Spectra of $\gamma$-unstable nuclei}\label{numgam}

Rms fits of spectra have been performed, using the quality measure 
\begin{equation}\label{eq:e99}
\sigma = \sqrt{ { \sum_{i=1}^n (E_i(exp)-E_i(th))^2 \over
(n-1)E(2_1^+)^2 } }.
\end{equation}
The theoretical predictions for the levels of the ground state band are obtained from Eq. (\ref{Egsb}),
while the levels of the quasi-$\beta_1$ band are obtained from Eq. (\ref{Eb1}).
The levels of the quasi-$\gamma_1$ band are obtained through their degeneracies to members of the ground state band,
mentioned below Eq. (\ref{eq:e7}). 

The results shown in Table 1 have been obtained for $\delta=\lambda=0$. (The Xe and Ba isotopes have already been considered
in Ref. \cite{first}.) 
One can easily verify that 
different choices for $\delta$ and $\lambda$ lead to a renormalization of the parameter values $a$ and $\beta_0$, 
the predicted energy levels remaining exactly the same. 

Concerning the physical content of the parameter $a$, it is instructive to consider in detail in Table~1 the Xe isotopes
(already discussed in Ref. \cite{first}), known \cite{Casten} to lie in a $\gamma$-unstable region. 
They extend from the borders of the neutron shell ($^{134}$Xe$_{80}$ is just below the N=82 shell closure)
to the midshell ($^{120}$Xe$_{66}$) and even beyond, exhibiting increasing collectivity (increasing $R_{4/2}=E(4_1^+)/E(2_1^+)$ ratios)
from the border to the mishell. 
Moving from the border of the neutron shell to the midshell, the following remarks apply

i) $^{134}$Xe and $^{132}$Xe are almost pure vibrators. Therefore no need for deformation dependence of the mass
exists, the least square fitting leading to $a=0$. Furthermore, no $\beta_0$ term is needed in the potential, 
the fitting therefore leading to $\beta_0=0$, {\it i.e.}, to pure harmonic behaviour. 

ii) In the next two isotopes ($^{130}$Xe and $^{128}$Xe) the need to depart from the pure harmonic oscillator 
becomes clear, the fitting leading therefore to nonzero $\beta_0$ values. However, there is still no need 
of dependence of the mass on the deformation, the fitting still leading to $a=0$. 

iii) Beyond $^{126}$Xe both the $\beta_0$ term in the potential and the deformation dependence of the mass become 
necessary, leading to nonzero values of both $\beta_0$ and $a$. 

Other chains of isotopes also show similar behavior.

\subsection{Spectra of axially symmetric deformed nuclei}\label{numdef}

Fits of spectra of deformed rare earth and actinide nuclei are shown in Table~2. 
The energy levels of the ground state band and the $\beta_1$ band (both having $n_\gamma=0$ and $K=0$) are 
obtained from Eqs. (\ref{Egsb}) and (\ref{Eb1}) respectively, while the levels of the $\gamma_1$ band are obtained 
from Eq. (\ref{Egsb}) with $n_\gamma=1$ and $K=2$. 
Again, the choice $\delta=\lambda=0$ has been made, and it is seen that 
different choices for $\delta$ and $\lambda$ lead to a renormalization of the parameter values $a$, $\beta_0$, and $c$,  
the predicted energy levels remaining exactly the same. 

The quality of the fits obtained can also be seen in Table~3, where the calculated energy levels 
of $^{162}$Dy and $^{238}$U are compared to experiment. 

The following remarks apply.

1) Both the bandheads and the spacings within bands are in general well reproduced.
This is particularly true for the ground state and the $\gamma_1$ bands.
The deviation in the gsb of $^{162}$Dy reaches 0.6\% at $L=18$, while in the gsb of $^{238}$U it reaches
1.7\% at $L=30$. The experimental levels of the $\gamma_1$ band of $^{162}$Dy (up to $L=14$) extend over 28.4 energy units, 
while the corresponding theoretical predictions spread over 28.7 units, the difference being of the order of 1\%.
Similarly in $^{238}$U the experimental spread of the $\gamma_1$ band (up to $L=27$) is 89.1 energy units, while the theoretical one is
87.3 units, the difference being of the order of 2\%.    

2) However we remark that the theoretical level spacings within the $\beta_1$ bands are larger than the experimental ones. 
This should be attributed to the shape of the Davidson potential, which raises to infinity at large $\beta$, 
pushing $\beta$ bands higher and increasing their interlevel spacing. It is known that this problem can be avoided
by using a potential going to some finite value at large $\beta$ \cite{finitew}, like the Morse potential \cite{Morse}. 

\subsection{$B(E2)$s of $\gamma$-unstable nuclei}\label{BE2gam}

$B(E2)$s within the ground state band, as well as interband $B(E2)$s for which experimental 
data exist for several nuclei, have been calculated using the procedure described in subsec. X.A~. 
The results are shown in Table 4, the overall agreement being good.

\subsection{$B(E2)$s of axially symmetric deformed nuclei}\label{BE2def}

$B(E2)$s within the ground state band, as well as interband $B(E2)$s for which experimental 
data exist for several nuclei, have been calculated using the procedure described in subsec. X.B~. 
The results are shown in Table 5. The overall agreement is good for transitions within the ground state band (gsb), 
as well as for transitions connecting the $\gamma_1$ band to the gsb, while transitions from the $\beta_1$ band to the gsb 
tend to be overpredicted. One should remember at this point that the $\beta_1$ band was the one giving 
poor results also in the case of the spectra, in subsec. XI.B~.    

\section{Connection to earlier work} 

It is instructive to examine the relation between the present approach and earlier numerical work. 

1) The formalism of subsection III.B clarifies the relation between the present approach and the numerical solution  
of Kumar and Baranger \cite{Kumar}, who used a matrix of the form (\ref{gmatr}) with
$g_{ij}$, $i,j=1,2,3$ the same as in Eq. (\ref{gmatrel}), but with   
\begin{equation}
g_{44}= B_{00}, \quad g_{55}= B_{2'2'}, \quad g_{45}=g_{54}=B_{02'},  
\end{equation}
where $B_{00}$, $B_{2'2'}$, $B_{02'}$, as well as the moments of inertia ${\cal J}_i$ ($i=1,2,3$)
and the potential $V$ have been treated as seven arbitrary functions of the variables 
$\beta_0=\beta \cos \gamma$ and $\beta_{2'}=\beta \sin \gamma$ [while in the Bohr formulation \cite{Bohr}
$a_0=\beta \cos \gamma$ and $a_2=\beta \sin\gamma /\sqrt{2}$ are used].  
On one hand, the present solution is a special case of Ref. \cite{Kumar}, since it contains no non-diagonal terms 
$g_{45}=g_{54}$.  
On the other hand, in the present approach the above mentioned quantities are interrelated by the overall symmetry 
in a specific way, greatly reducing the number of free parameters (down to two or three in total). 
It should be pointed out that the functional dependence of the mass 
on the deformation for the potential used is dictated by SUSYQM.
Therefore, the successful prediction of the 
behavior of the moments of inertia, for example, provides credit for the present approach.
What we see, independently of the parameter values, is that the increase of the moments of inertia
as a function of deformation is moderated by the $f^2$ factor, which can be seen as a result 
of the dependence of the mass on the deformation, or, alternatively, as seen in subsection III.B, 
as a result of using a curved space. 

2) It should be pointed out that in Ref. \cite{QT4267} the equivalence between the position dependent mass case 
and the curved space approach has been established in the special case of $\kappa=2$ and $\delta=\lambda=0$
(see Eq. (\ref{Hdelta}) for the meaning of the symbols), which represents the BenDaniel and Duke Hamiltonian \cite{BenD}
\begin{equation}
H_{BD} = -{\hbar^2 \over 2} \nabla f^2 \nabla  + V_{BD}. 
\end{equation}  
This resembles the collective Hamiltonian
\begin{equation}
H_{coll}= -{\hbar^2 \over 2} \Sigma_{i,j} {\partial \over \partial q_i} [{\cal M}(q)_{ij}]^{-1} {\partial \over \partial q_j} +V(q) 
\end{equation}
used by Libert et al. \cite{Libert} in mean field calculations, in which a tensor mass appears. 

\section{Conclusion}

In the present work analytical solutions are obtained for a Bohr Hamiltonian in which
the mass has been allowed to depend on the deformation.

From the mathematical point of view, this is achieved through the use of techniques of supersymmetric quantum 
mechanics \cite{PR,SUSYQM}, involving a deformed shape invariance condition \cite{Q2929}. Analytical expressions 
for the spectra and wave functions have been obtained.

From the physics point of view, spectra and $B(E2)$ transition rates have been calculated for $\gamma$-unstable,
axially symmetric prolate deformed, and triaxial nuclei, implementing the usual approximations in each case,
and compared to experimental data for the first two cases.
The main new result is that the dependence of the mass on the deformation moderates the increase of the moment
of inertia with the deformation, removing an important drawback \cite{Ring} of the model.
It should be emphasized that the functional dependence of the mass on the deformation for the potential used 
is dictated by SUSYQM, thus the correction in the behavior of the moments of inertia is a general effect,
independent of any specific parameter value combinations.

However, certain discrepancies with experimental data remain, especially related to the $\beta_1$-band and its 
interband transitions. It should be remembered at this point that in the present study separation of variables 
has been achieved by {\sl assuming} that the potential either is independent of the $\gamma$-variable, or it has 
the exactly separable form of Eq. (\ref{eq:e7b}). Furthermore, the approximations related to Eqs. (\ref{Qdef}) and (\ref{Qtri})
have been implemented. 
Recently, the numerical solution of the Bohr Hamiltonian for any value 
of $\beta$ and $\gamma$, avoiding all these approximations,
 has been achieved in the framework of the powerful algebraic collective model \cite{Rowe,Turner,Welsh}. The detailed study 
of discrepancies from experimental data both in the SUSYQM framework and in the context of the algebraic model,
especially for multi-phonon excitations \cite{Caprio2}, could shed light on the origins of these discrepancies. 

As it has already been mentioned, the form of the dependence of the mass on the deformation is dictated by SUSYQM for the potential used 
in the $\beta$ degree of freedom. In the present work, the Davidson potential has been used, called the Deformation Dependent Mass (DDM) 
Davidson model.  
The application of the SUSYQM approach to the Bohr Hamiltonian with the Kratzer potential \cite{FV1,FV2} is receiving attention.

\section*{Acknowledgements}

The authors are thankful to F. Iachello for suggesting the project and for useful discussions.
One of authors (N. M.) acknowledges the support of the Bulgarian Scientific Fund under contract DID-02/16-17.12.2009.

\section*{Appendix 1}

When using Eq. (\ref{normaliz}) in numerical calculations, problems can appear 
because of $\Gamma(x)$ functions with large $x$. These problems can be avoided by using 
Eq. 6.1.16 of Ref. \cite{AbrSte} 
\begin{equation}
\Gamma (n+z)= (n-1+z) (n-2+z) \dots (1+z) \Gamma(1+z) .
\end{equation}
In the normalization factors we need the ratio of 
\begin{equation}
\Gamma\left(n + {\Delta_1+\Delta_2 \over 2} +1   \right)
\end{equation}
over
\begin{equation}
\Gamma\left(n + {\Delta_2 \over 2} +1   \right).
\end{equation}
Let us call $I$ the integer part of $\Delta_2/2$ and $r$ the rest of it, i.e., 
\begin{equation}
{\Delta_2 \over 2} = I+r.
\end{equation}
Then we have 
\begin{equation}
\begin{split}
& \Gamma\left(n + {\Delta_1+\Delta_2 \over 2} +1   \right)= \Gamma\left(I + (n + \Delta_1/2+r+1)   \right)  \\
& = (I-1+n+\Delta_1/2+r+1) \\
& (I-2+n+\Delta_1/2+r+1) \dots  (1+n+\Delta_1/2+r+1) \\
& \Gamma(1+n + \Delta_1/2+r+1), 
\end{split}
\end{equation}
\begin{equation}
\begin{split}
& \Gamma\left(n + {\Delta_2 \over 2} +1   \right)= \Gamma\left(I + (n +r+1)   \right) \\
& = (I-1+n+r+1) (I-2+n+r+1) \dots \\
& (1+n+r+1) \Gamma(1+n +r+1). 
\end{split} 
\end{equation}
Their ratio becomes 
\begin{equation}
\begin{split}
& {\Gamma\left(n + {\Delta_1+\Delta_2 \over 2} +1   \right) \over 
\Gamma\left(n + {\Delta_2 \over 2} +1   \right)} =
{(I-1+n+\Delta_1/2+r+1)\over (I-1+n+r+1)} \\
&   {(I-2+n+\Delta_1/2+r+1) \over (I-2+n+r+1)} \dots  
{(1+n+\Delta_1/2+r+1) \over (1+n+r+1)} \\
& {\Gamma(1+n + \Delta_1/2+r+1)\over \Gamma(1+n +r+1)} \\
& = \left(1+ {\Delta_1/2\over (I-1+n+r+1)} \right)  \\
& \left(1+ {\Delta_1/2\over (I-2+n+r+1)} \right) \dots \left(1+ {\Delta_1/2\over (1+n+r+1)} \right) \\
& {\Gamma(1+n + \Delta_1/2+r+1)\over \Gamma(1+n +r+1)},
\end{split}
\end{equation}
in which one does not have to calculate  $\Gamma(x)$ functions with large $x$. The only large numbers appear 
in denominators of fractions accompanying 1, which do not pose any problem.

\newpage

\begin{table*}

\caption{Comparison of theoretical predictions of the 
$\gamma$-unstable Bohr Hamiltonian with $\beta$-dependent mass 
(with $\delta =\lambda =0$)
to experimental data \cite{NDS} of
rare earth and actinide nuclei with $R_{4/2} \leq 2.6$ and known $0_2^+$
and $2_{\gamma}^+$ states.  The $R_{4/2}=E(4_1^+)/E(2_1^+)$
ratios, as well as the $\beta$ and $\gamma$ bandheads, normalized
to the $2_1^+$ state and labelled by
$R_{0/2}=E(0_{\beta}^+)/E(2_1^+)$ and
$R_{2/2}=E(2_{\gamma}^+)/E(2_1^+)$ respectively, are shown. 
$\beta_0$ and $a$ are free parameters, related to the Davidson potential 
[Eq. (\ref{eq:e16})] and to the dependence of the mass on the deformation [Eq. (\ref{eq:e17})].
The angular momenta of the highest levels of the ground state, $\beta$
and $\gamma$ bands included in the rms fit are labelled by $L_g$,
$L_\beta$, and $L_\gamma$ respectively, while $n$ indicates the
total number of levels involved in the fit and $\sigma$ is the
quality measure of Eq. (\ref{eq:e99}). The theoretical predictions are obtained 
from the formulae mentioned below Eq. (\ref{eq:e99}).
The Xe and Ba isotopes have already been considered
in Ref. \cite{first}.
See subsec. \ref{numgam} for further discussion. }

\bigskip

\begin{tabular}{ r r r r  r r r r  r r r r r r}
\hline nucleus & $R_{4/2}$ & $R_{4/2}$ & $R_{0/2}$& $R_{0/2}$
&$R_{2/2}$ & $R_{2/2}$ & $\beta_0$ & $a$ &
$L_g$ & $L_\beta$ & $L_\gamma$ & $n$ & $\sigma$ \\
        & exp &  th  & exp & th  & exp & th &  &  &  &  &  &   &  \\

\hline

$^{98}$Ru  & 2.14 & 2.14 & 2.0 &  2.4 &  2.2 &  2.1 & 0.99 & 0.020 
& 24 &   0 &   4 & 15 & 0.277 \\
$^{100}$Ru & 2.27 & 2.24 & 2.1 &  2.7 &  2.5 &  2.2 & 1.19 & 0.048
& 28 &   0 &   4 & 17 & 0.315 \\
$^{102}$Ru & 2.33 & 2.20 & 2.0 &  2.4 &  2.3 &  2.2 & 1.05 & 0.059
& 16 &   0 &   5 & 12 & 0.364 \\
$^{104}$Ru & 2.48 & 2.34 & 2.8 &  3.0 &  2.5 &  2.3 & 1.40 & 0.083
&  8 &  2  &   8 & 12 & 0.429 \\ 

$^{102}$Pd & 2.29 & 2.24 & 2.9 & 2.3 & 2.8 & 2.2 & 1.08 & 0.081
& 26 &  4  &  4 & 18 & 0.326 \\
$^{104}$Pd & 2.38 & 2.21 & 2.4 & 2.6 & 2.4 & 2.2 & 1.15 & 0.034
& 18 &  2  &  4 & 13 & 0.397 \\
$^{106}$Pd & 2.40 & 2.16 & 2.2 & 2.2 & 2.2 & 2.2 & 0.91 & 0.062
& 16 &  4  &  5 & 14 & 0.409 \\
$^{108}$Pd & 2.42 & 2.26 & 2.4 & 2.3 & 2.1 & 2.3 & 1.09 & 0.103
& 14 &  4  &  4 & 12 & 0.318 \\
$^{110}$Pd & 2.46 & 2.31 & 2.5 & 2.0 & 2.2 & 2.3 & 0.99 & 0.195 
& 12 &  10 &  4 & 14 & 0.354 \\
$^{112}$Pd & 2.53 & 2.29 & 2.6 & 2.5 & 2.1 & 2.3 & 1.21 & 0.086 
&  6 &   0 &  3 &  5 & 0.485 \\
$^{114}$Pd & 2.56 & 2.31 & 2.6 & 2.8 & 2.1 & 2.3 & 1.30 & 0.076 
& 16 &   0 & 11 & 18 & 0.722 \\
$^{116}$Pd & 2.58 & 2.36 & 3.3 & 3.4 & 2.2 & 2.4 & 1.52 & 0.062 
& 16 &   0 &  9 & 16 & 0.609 \\

$^{106}$Cd & 2.36 & 2.25 & 2.8 & 2.9 & 2.7 & 2.3 & 1.28 & 0.028 
& 12 & 0 & 2 &  7 & 0.268 \\
$^{108}$Cd & 2.38 & 2.14 & 2.7 & 2.2 & 2.5 & 2.1 & 0.91 & 0.041 
& 24 & 0 & 5 & 16 & 0.528 \\
$^{110}$Cd & 2.35 & 2.08 & 2.2 & 1.9 & 2.2 & 2.1 & 0.00 & 0.061 
& 16 & 6 & 5 & 15 & 0.415 \\
$^{112}$Cd & 2.29 & 2.05 & 2.0 & 1.9 & 2.1 & 2.0 & 0.00 & 0.033 
& 12 & 8 &11 & 20 & 0.523 \\
$^{114}$Cd & 2.30 & 2.06 & 2.0 & 1.9 & 2.2 & 2.1 & 0.00 & 0.041
& 14 & 4 & 3 & 11 & 0.418 \\
$^{116}$Cd & 2.38 & 2.16 & 2.5 & 2.7 & 2.4 & 2.2 & 1.14 & 0.000
& 14 & 2 & 3 & 10 & 0.387 \\ 
$^{118}$Cd & 2.39 & 2.19 & 2.6 & 2.9 & 2.6 & 2.2 & 1.21 & 0.002
& 14 & 0 & 3 &  9 & 0.429 \\
$^{120}$Cd & 2.38 & 2.20 & 2.7 & 2.9 & 2.6 & 2.2 & 1.22 & 0.006
& 16 & 0 & 2 &  9 & 0.412 \\

$^{118}$Xe & 2.40 & 2.32 & 2.5 & 2.6 & 2.8 & 2.3 & 1.27 & 0.103
& 16 & 4 &10 & 19 & 0.319 \\
$^{120}$Xe & 2.47 & 2.36 & 2.8 & 3.4 & 2.7 & 2.4 & 1.51 & 0.063
& 26 & 4 & 9 & 23 & 0.524 \\
$^{122}$Xe & 2.50 & 2.40 & 3.5 & 3.3 & 2.5 & 2.4 & 1.57 & 0.096 
& 16 & 0 & 9 & 16 & 0.638 \\
$^{124}$Xe & 2.48 & 2.36 & 3.6 & 3.5 & 2.4 & 2.4 & 1.55 & 0.051 
& 20 & 2 &11 & 21 & 0.554 \\
$^{126}$Xe & 2.42 & 2.33 & 3.4 & 3.1 & 2.3 & 2.3 & 1.42 & 0.064 
& 12 & 4 & 9 & 16 & 0.584 \\
$^{128}$Xe & 2.33 & 2.27 & 3.6 & 3.5 & 2.2 & 2.3 & 1.42 & 0.000 
& 10 & 2 & 7 & 12 & 0.431 \\
$^{130}$Xe & 2.25 & 2.21 & 3.3 & 3.1 & 2.1 & 2.2 & 1.27 & 0.000 
& 14 & 0 & 5 & 11 & 0.347 \\
$^{132}$Xe & 2.16 & 2.00 & 2.8 & 2.0 & 1.9 & 2.0 & 0.00 & 0.000 
&  6 & 0 & 5 &  7 & 0.467 \\
$^{134}$Xe & 2.04 & 2.00 & 1.9 & 2.0 & 1.9 & 2.0 & 0.00 & 0.000 
&  6 & 0 & 5 &  7 & 0.685 \\

$^{130}$Ba & 2.52 & 2.42 & 3.3 & 3.2 & 2.5 & 2.4 & 1.60 & 0.118 
& 12 & 0 & 6 & 11 & 0.352 \\  
$^{132}$Ba & 2.43 & 2.29 & 3.2 & 2.8 & 2.2 & 2.3 & 1.29 & 0.059 
& 14 & 0 & 8 & 14 & 0.619 \\
$^{134}$Ba & 2.32 & 2.16 & 2.9 & 2.7 & 1.9 & 2.2 & 1.12 & 0.000 
&  8 & 0 & 4 &  7 & 0.332 \\
$^{136}$Ba & 2.28 & 2.00 & 1.9 & 2.0 & 1.9 & 2.0 & 0.00 & 0.000 
&  6 & 0 & 2 &  4 & 0.250 \\  
$^{142}$Ba & 2.32 & 2.38 & 4.3 & 4.3 & 4.0 & 2.4 & 1.72 & 0.028 
& 14 & 0 & 2 &  8 & 0.609 \\

$^{134}$Ce & 2.56 & 2.34 &  3.7 &  3.9 &  2.4 & 2.3 &  1.59 & 0.019 &
34 & 2 & 8 & 25 & 0.527 \\
$^{136}$Ce & 2.38 & 2.11 &  1.9 &  2.1 &  2.0 & 2.1 &  0.82 & 0.034 &
16 & 0 & 3 & 10 & 0.457 \\
$^{138}$Ce & 2.32 & 2.00 &  1.9 &  2.0 &  1.9 & 2.0 &  0.00 & 0.000 &
14 & 0 & 2 &  8 & 0.314 \\

$^{140}$Nd & 2.33 & 2.05 &  1.8 &  1.9 & 1.9 & 2.1 & 0.00 & 0.037 &
 6 & 0 & 2 &  4 & 0.192 \\
$^{148}$Nd & 2.49 & 2.36 &  3.0 &  2.8 & 4.1 & 2.4 & 1.38 & 0.110 &
12 & 8 & 4 & 13 & 0.764 \\

$^{140}$Sm & 2.35 & 2.29 & 1.9 & 1.9 & 2.7 & 2.3 & 0.92 & 0.196 &
 8 & 0 & 2 &  5 & 0.207 \\
$^{142}$Sm & 2.33 & 2.06 & 1.9 & 1.9 & 2.2 & 2.1 & 0.33 & 0.044 &
 8 & 0 & 2 &  5 & 0.147 \\

$^{142}$Gd & 2.35 & 2.21 & 2.7 & 2.8 & 1.9 & 2.2 & 1.20 & 0.020 &
16 & 0 & 2 &  9 & 0.231 \\
$^{144}$Gd & 2.35 & 2.33 & 2.5 & 2.5 & 2.5 & 2.3 & 1.26 & 0.112 &
 6 & 0 & 2 &  4 & 0.124 \\
$^{152}$Gd & 2.19 & 2.13 & 1.8 & 1.8 & 3.2 & 2.1 & 0.00 & 0.104 &
16 & 10& 7 & 19 & 0.635 \\
 
$^{154}$Dy  & 2.23 & 2.15 & 2.0 & 2.0 & 3.1 & 2.1 & 0.75 & 0.083 &
26 & 10 & 7 & 24 & 0.530 \\

$^{156}$Er  & 2.32 & 2.25 & 2.7 & 2.8 & 2.7 & 2.3 & 1.24 & 0.043 &
20 & 4 & 5 & 16 & 0.450 \\

\hline
\end{tabular}
\end{table*}

\begin{table*}
\setcounter{table}{0} \caption{(continued)}

\bigskip

\begin{tabular}{ r r r r  r r r r  r r r r r r}
\hline nucleus & $R_{4/2}$ & $R_{4/2}$ & $R_{0/2}$& $R_{0/2}$
&$R_{2/2}$ & $R_{2/2}$ & $\beta_0$ & $a$ & 
$L_g$ & $L_\beta$ & $L_\gamma$ & $n$ & $\sigma$ \\
        & exp &  th  & exp & th  & exp & th &  &  &  &  &    & &  \\

\hline

$^{186}$Pt & 2.56 & 2.42 & 2.5 & 3.7 & 3.2 & 2.4 & 1.71 & 0.085 &
26 & 6 & 10 & 25 & 0.813 \\ 
$^{188}$Pt & 2.53 & 2.37 & 3.0 & 3.3 & 2.3 & 2.4 & 1.52 & 0.076 &
16 & 2 & 4 & 12 & 0.637 \\
$^{190}$Pt & 2.49 & 2.28 & 3.1 & 3.4 & 2.0 & 2.3 & 1.42 & 0.015 &
18 & 2 & 6 & 15 & 0.637 \\
$^{192}$Pt & 2.48 & 2.34 & 3.8 & 3.7 & 1.9 & 2.3 & 1.56 & 0.032 &
10 & 0 & 8 & 12 & 0.681 \\
$^{194}$Pt & 2.47 & 2.36 & 3.9 & 3.6 & 1.9 & 2.4 & 1.55 & 0.049 &
10& 4 & 5 & 11 & 0.667 \\
$^{196}$Pt & 2.47 & 2.33 & 3.2 & 2.9 & 1.9 & 2.3 & 1.37 & 0.079 &
10 & 2 & 6 & 11 & 0.639 \\ 
$^{198}$Pt & 2.42 & 2.21 & 2.2 & 2.2 & 1.9 & 2.2 & 0.96 & 0.089 &
 6 & 2 & 4 & 7  & 0.370 \\
$^{200}$Pt & 2.35 & 2.00 & 2.4 & 2.0 & 1.8 & 2.0 & 0.00 & 0.000 &
 4 & 0 & 4 &  5 & 0.392 \\

\hline
\end{tabular}
\end{table*}

\newpage

\begin{table*}

\caption{Same as Table~1, but for axially symmetric prolate deformed 
rare earth and actinide nuclei with $R_{4/2}$ $>$ 2.9~.
$\beta_0$, $a$, and $c$ are free parameters, related to the Davidson potential 
[Eq. (\ref{eq:e16})],  to the dependence of the mass on the deformation [Eq. (\ref{eq:e17})],
and to the $\gamma$-potential [Eq. (\ref{wgamma})]. 
The theoretical predictions are obtained from the equations mentioned in  
subsec. \ref{numdef}, where further
discussion can be found. }

\bigskip

\begin{tabular}{ l r r r  r r r r r r r r r r r}
\hline nucleus & $R_{4/2}$ & $R_{4/2}$ & $R_{0/2}$& $R_{0/2}$
&$R_{2/2}$ & $R_{2/2}$ & $\beta_0$ & $c$   & $a$ &
$L_g$ & $L_\beta$ & $L_\gamma$ & $n$ & $\sigma$ \\
        & exp &  th  & exp & th  & exp & th &  &  &  & & &  &   &  \\

\hline

$^{150}$Nd & 2.93 & 3.13 & 5.2 &  7.9 &  8.2 &  5.8 &   0.0 & 2.1 &  0.003 
& 14 &   6 &   4 & 13 & 2.012 \\
 
$^{152}$Sm & 3.01 & 3.14 & 5.6 &  8.4 &  8.9 &  6.5 &   0.0 & 2.4 & 0.000
& 16 &  14 &   9 & 23 & 3.327 \\
$^{154}$Sm & 3.25 & 3.27 & 13.4 & 13.0 & 17.6 & 18.6 & 1.30 & 6.9 & 0.021
& 16 &   6 &   7 & 17 & 0.515 \\

$^{154}$Gd & 3.02 & 3.09 &  5.5 &  6.5 &  8.1 &  4.1 & 0.0  & 1.4 & 0.024
& 26 & 26  &   7 & 32 & 3.546 \\ 
$^{156}$Gd & 3.24 & 3.25 & 11.8 & 10.8 & 13.0 & 14.3 & 0.0  & 5.3 & 0.026
& 26 & 12  &   16 & 34 & 0.933 \\
$^{158}$Gd & 3.29 & 3.29 & 15.0 & 14.5 & 14.9 & 15.1 & 1.99 & 5.3 & 0.025
& 12 &  6  &   6 & 14 & 0.323 \\
$^{160}$Gd & 3.30 & 3.30 & 17.6 & 17.3 & 13.1 & 13.2 & 2.38 & 4.5 & 0.020
& 16 &  4  &   8 & 17 & 0.125 \\
$^{162}$Gd & 3.29 & 3.30 & 19.8 & 19.8 & 12.0 & 12.1 & 2.52 & 4.1 & 0.008
& 14 &  0  &   4 & 10 & 0.078 \\

$^{156}$Dy & 2.93 & 3.13 &  4.9 &  7.4 &  6.5 &  5.3 & 0.0  & 1.9 & 0.014 & 
28 &  10 &  13 & 31 & 1.789 \\
$^{158}$Dy & 3.21 & 3.22 & 10.0 &  9.6 &  9.6 & 10.3 & 0.26  & 3.8 & 0.023 & 
28 &  8 &  8 & 25 & 0.496 \\
$^{160}$Dy & 3.27 & 3.27 & 14.7 & 14.7 & 11.1 & 12.1 & 1.92 & 4.3 & 0.005 & 
28 & 4 & 23 & 38 & 0.510 \\
$^{162}$Dy & 3.29 & 3.30 & 17.3 & 15.7 & 11.0 & 11.2 & 2.23 & 3.8 & 0.020 &
18 &  8 & 14 & 26 & 0.742 \\
$^{164}$Dy & 3.30 & 3.30 & 22.6 & 22.5 & 10.4 & 10.2 & 2.68 & 3.4 & 0.000 & 
20 & 0 & 10 & 19 & 0.100 \\
$^{166}$Dy & 3.31 & 3.31 & 15.0 & 14.9 & 11.2 & 11.2 & 2.39 & 3.7 & 0.047 & 
6 & 2 & 5 & 8 & 0.077 \\

$^{160}$Er & 3.10 & 3.16 &  7.1 &  8.1 &  6.8 & 6.6 & 0.00 &  2.4 & 0.013
& 26 & 2 & 5 & 18 & 0.699\\
$^{162}$Er & 3.23 & 3.23 & 10.7  & 10.7 &  8.8 & 10.1 & 1.29 & 3.7 & 0.013
&20 & 4 & 12 & 23 & 0.770 \\ 
$^{164}$Er & 3.28 & 3.27 & 13.6 & 12.2 &  9.4 &  9.6 & 1.83 & 3.3 & 0.026
&22 & 10 & 18 & 33 & 0.918 \\
$^{166}$Er & 3.29 & 3.28 & 18.1 & 16.8 &  9.8 &  9.9 & 2.22 &  3.4 & 0.002
& 16 & 10 & 14 & 26 & 0.698 \\
$^{168}$Er & 3.31 & 3.31 & 15.3 & 14.4 & 10.3 & 10.2 & 2.29 & 3.4 & 0.041
& 18 & 6 & 8 & 19 & 0.404 \\
$^{170}$Er & 3.31 & 3.30 & 11.3 & 10.1 & 11.9 & 12.9 & 1.64 & 4.4 & 0.083
&24 & 10 & 19 & 35 & 0.837 \\

$^{162}$Yb & 2.92 & 3.07 & 3.6 & 6.8 &   4.8 & 4.0 & 0.00 &  1.4 & 0.003 &
24 & 0 & 4 & 15 & 1.036\\
$^{164}$Yb & 3.13 & 3.18 & 7.9 & 8.3 &   7.0 & 7.4 & 0.00 &  2.7 & 0.023 &
18 & 0 & 5 & 13 & 0.357\\
$^{166}$Yb & 3.23& 3.23 & 10.2 & 8.9&    9.1 & 9.7 & 0.66 &  3.5 & 0.038 &
24 & 10 & 13 & 29 & 0.973 \\
$^{168}$Yb & 3.27 & 3.26 & 13.2& 11.2 & 11.2 &11.5 & 1.52 & 4.1 & 0.028 &
34 & 4 & 7 & 25 & 1.070 \\
$^{170}$Yb & 3.29 & 3.27& 12.7 & 11.2 & 13.6 &14.1 & 1.36 & 5.1 & 0.035 &
20 & 10 & 17 & 31 & 0.963 \\
$^{172}$Yb & 3.31 & 3.30 & 13.2 & 12.2& 18.6 & 18.9 & 1.66  & 6.6 & 0.055 &
16 & 10 & 5 & 17 & 0.742 \\
$^{174}$Yb & 3.31 & 3.31 & 19.4 & 19.3& 21.4 & 21.5 & 2.44 & 7.5 & 0.019 &
20 & 4 & 5 & 16 & 0.104 \\
$^{176}$Yb & 3.31 & 3.30& 13.9 &  13.7& 15.4 & 15.5 & 1.97 & 5.4 & 0.036 &
20 & 2 & 5 & 15 & 0.287 \\  
$^{178}$Yb & 3.31 & 3.27& 15.7 &  15.5& 14.5 & 14.6 & 1.88 & 5.3 & 0.000 & 
6 & 4 & 2 & 6 & 0.127 \\

$^{166}$Hf & 2.97 & 3.08 &  4.4 &  6.9 &  5.1 & 4.3 &  0.00 & 1.5 & 0.006 &
22 & 0 & 3 & 13 & 0.873\\
$^{168}$Hf & 3.11 & 3.17 &  7.6 &  8.1 &  7.1 & 6.9 &  0.00 & 2.5 & 0.023 &
22 & 4 & 4 & 16 & 0.494\\
$^{170}$Hf & 3.19 & 3.21 &  8.7 &  8.7 &  9.5 & 8.8 &  0.00 & 3.2 & 0.033 &
34 & 4 & 4 & 22& 0.970\\
$^{172}$Hf & 3.25 & 3.24 &  9.2 &  9.8 & 11.3 & 11.7 & 0.00  & 4.3 & 0.031 &
38 & 4 & 6 & 26 & 0.549 \\
$^{174}$Hf & 3.27 & 3.25 &  9.1 & 10.4 & 13.5 & 13.6 & 0.00 & 5.0 & 0.033 &
26 &  4 & 5 & 19 & 0.832 \\
$^{176}$Hf & 3.28 & 3.28 & 13.0 & 11.5 & 15.2 & 16.1 & 1.31 & 5.8 & 0.038 &
18 & 10 & 8 & 21 & 0.950 \\
$^{178}$Hf & 3.29 & 3.28 & 12.9 & 12.3 & 12.6 & 13.0 & 1.70 & 4.6 & 0.028 &
18 & 6 & 6 & 17 & 0.356 \\
$^{180}$Hf & 3.31 & 3.30 & 11.8 & 11.5 & 12.9 & 13.0 & 1.92 & 4.4 & 0.068 &
12 & 4 & 5 & 12 & 0.157 \\

$^{176}$W  & 3.22 & 3.21 & 7.8 &  9.1 &  9.6 &  9.5 & 0.00 &  3.5 & 0.027 &
22& 4 & 5 & 17 & 0.881\\
$^{178}$W  & 3.24 & 3.22&  9.4 &  8.6 & 10.5 &  8.9 & 0.00 &  3.2 & 0.039 &
 18 & 10 & 2 & 15 & 0.987 \\
$^{180}$W  & 3.26 & 3.25&  14.6& 13.1 & 10.8 & 11.5 & 1.64 &  4.2 & 0.000 &
24 & 0 & 7 & 18 & 0.603 \\
$^{182}$W  & 3.29 & 3.29 & 11.3& 11.5 & 12.2 & 12.5 & 1.77 &  4.3 & 0.050 &
18 & 4 & 6 & 16 & 0.195 \\
$^{184}$W  & 3.27 & 3.28 & 9.0 &  8.9 &  8.1 &  8.0 & 1.57 &  2.7 & 0.080 &
 10 & 4 & 6 & 12 & 0.093 \\ 
$^{186}$W  & 3.23 & 3.25&  7.2 &  7.2 &  6.0 &  6.3 & 1.20 &  2.1 & 0.099 &
14 & 4 & 6 & 14 & 0.130 \\

$^{176}$Os & 2.93 & 3.10 &  4.5 & 6.9 & 6.4 & 4.6 & 0.00 & 1.6 & 0.016 &
24& 6 & 5 & 19 & 1.747 \\
$^{178}$Os & 3.02 & 3.12 &  4.9 & 7.2 & 6.6 & 5.1 & 0.00 & 1.8 & 0.017 &
16& 6 & 5 & 15 & 1.836 \\
$^{180}$Os & 3.09 & 3.22 &  5.6 & 7.1 & 6.6 & 6.9 & 0.00 & 2.4 & 0.078 &
10& 6 & 7 & 14 & 1.021 \\
$^{184}$Os & 3.20 & 3.21 &  8.7 & 9.9 &  7.9 & 8.5 & 1.21 & 3.1 & 0.011 &
22 & 0 & 6 & 16 & 0.886 \\ 
$^{186}$Os & 3.17 & 3.19 &  7.7 & 7.0 &  5.6 & 6.0 & 0.00 & 2.1 & 0.063 &
14 & 10 & 13 & 24 & 0.702 \\
$^{188}$Os & 3.08 & 3.15 &  7.0 & 7.2 &  4.1 & 4.4 & 1.07 & 1.5 & 0.033 &
12 & 2 & 7 & 13 & 0.170 \\
$^{190}$Os & 2.93 & 3.07 &  4.9 & 5.6 &  3.0 & 3.1 & 0.00 & 1.0 & 0.051 &
10 & 2 & 6 & 11 & 0.419 \\

\hline
\end{tabular}
\end{table*}

\begin{table*}
\setcounter{table}{1} \caption{(continued)}

\bigskip

\begin{tabular}{ l r r r  r r r r r r r r r r r}
\hline nucleus & $R_{4/2}$ & $R_{4/2}$ & $R_{0/2}$& $R_{0/2}$
&$R_{2/2}$ & $R_{2/2}$ & $\beta_0$ & $c$   & $a$ & 
$L_g$ & $L_\beta$ & $L_\gamma$ & $n$ & $\sigma$ \\
        & exp &  th  & exp & th  & exp & th &  &  &  &  &  &  & &  \\

\hline

$^{228}$Ra & 3.21 & 3.24 & 11.3 & 11.0 & 13.3 & 13.3 & 0.57 & 5.0 & 0.016 
& 22 & 4 & 3 & 15 & 0.177 \\

$^{228}$Th & 3.24 & 3.26 & 14.4 & 14.3 & 16.8 & 17.0 & 1.50 & 6.4 & 0.002 
& 18 & 2 & 5 & 14 & 0.214 \\
$^{230}$Th & 3.27 & 3.27 & 11.9 & 11.6& 14.7 &  14.7 & 1.44  & 5.3 & 0.034
& 24 & 4 & 4 & 17 & 0.243 \\
$^{232}$Th & 3.28 & 3.28 & 14.8 & 14.0 & 15.9 & 16.5 & 1.80 & 5.9 & 0.022
& 30 & 10 & 12 & 31 & 0.426 \\
 
$^{232}$U  & 3.29 & 3.29 & 14.5 & 13.8 & 18.2 & 18.4 & 1.74 & 6.6 & 0.028
&20 & 10 & 4 & 18 & 0.394 \\
$^{234}$U  & 3.30 & 3.30 & 18.6 & 18.3 & 21.3 & 21.8 & 2.19 & 7.8 & 0.011
& 28 & 8 & 7 & 24 & 0.244 \\
$^{236}$U  & 3.30 & 3.30 & 20.3& 20.0 & 21.2 & 21.2 & 2.38 & 7.5 & 0.009
& 30 & 4 & 5 & 21 & 0.143 \\
$^{238}$U  & 3.30 & 3.31 & 20.6 & 20.6 & 23.6 & 24.7 & 2.38 & 8.8 & 0.009
& 30 & 4 & 27 & 43 & 0.665 \\

$^{238}$Pu & 3.31 & 3.31 & 21.4 & 21.4 & 23.3 & 23.3 & 2.61 & 8.1 &  0.016
&26 & 2 & 4 & 17 & 0.067 \\
$^{240}$Pu & 3.31 & 3.31 & 20.1 & 19.9 & 26.6 & 26.6 & 2.40 & 9.4 & 0.018
& 26 & 4 & 4 & 18 & 0.117 \\
$^{242}$Pu & 3.31 & 3.31 & 21.5 & 21.4 & 24.7 & 24.7 & 2.52 & 8.7 & 0.012
& 26 & 2 & 2 & 15 & 0.107 \\

$^{248}$Cm & 3.31 & 3.31& 25.0 & 24.8 & 24.2 & 24.3 & 2.72 & 8.5 & 0.004
&28 & 4 & 2 & 17 & 0.159 \\

$^{250}$Cf & 3.32 & 3.31& 27.0 & 26.9 & 24.2 & 24.2 & 2.88 & 8.4 & 0.003
&8 & 2 & 4 & 8 & 0.053 \\

\hline
\end{tabular}
\end{table*}

\newpage

\begin{table}

\caption{Normalized [to the energy of the first excited state, $E(2_1^+)$] energy levels of the ground state band (gsb) 
and the  $\beta_1$ and $\gamma_1$ bands of $^{162}$Dy and $^{238}$U,
obtained from the Bohr Hamiltonian with $\beta$-dependent mass for axially symmetric prolate deformed nuclei
using the parameters given in Table~2, compared to experimental data \cite{NDS}. See subsec. \ref{numdef}
for further discussion. }

\bigskip

\begin{tabular}{  r r r r r   r r r r r}
\hline 
  & $^{162}$Dy & $^{162}$Dy & $^{238}$U & $^{238}$U &    & $^{162}$Dy & $^{162}$Dy & $^{238}$U & $^{238}$U \\ 
L &  exp       & th         & exp       & th        &  L  &   exp      &   th       & exp       & th        \\      

\hline
   &  gsb  &  gsb  &  gsb  &  gsb  &    &$\gamma_1$ & $\gamma_1$ & $\gamma_1$ & $\gamma_1$ \\
 0 &  0.00 &  0.00 &  0.00 &  0.00 &  2 & 11.0 & 11.2 & 23.6 & 24.7 \\
 2 &  1.00 &  1.00 &  1.00 &  1.00 &  3 & 11.9 & 12.1 & 24.6 & 25.5 \\
 4 &  3.29 &  3.30 &  3.30 &  3.31 &  4 & 13.2 & 13.3 & 25.9 & 26.7 \\
 6 &  6.80 &  6.80 &  6.84 &  6.86 &  5 & 14.7 & 14.7 & 27.4 & 28.1 \\
 8 & 11.41 & 11.41 & 11.54 & 11.57 &  6 & 16.4 & 16.5 & 29.2 & 29.8 \\
10 & 17.04 & 17.01 & 17.27 & 17.33 &  7 & 18.5 & 18.5 & 31.2 & 31.7 \\
12 & 23.57 & 23.49 & 23.97 & 24.06 &  8 & 20.7 & 20.8 & 33.5 & 33.9 \\
14 & 30.90 & 30.74 & 31.51 & 31.63 &  9 & 23.3 & 23.3 & 36.0 & 36.3 \\
16 & 38.90 & 38.70 & 39.82 & 39.97 & 10 & 25.9 & 26.0 & 38.8 & 39.0 \\
18 & 47.58 & 47.28 & 48.78 & 48.98 & 11 & 29.0 & 28.9 & 41.7 & 41.9 \\
20 &       &       & 58.31 & 58.61 & 12 & 31.4 & 32.1 & 44.9 & 45.0 \\
22 &       &       & 68.31 & 68.77 & 13 & 35.5 & 35.5 & 48.3 & 48.3 \\
24 &       &       & 78.71 & 79.44 & 14 & 39.4 & 39.9 & 51.9 & 51.8 \\
26 &       &       & 89.46 & 90.55 & 15 &      &      & 55.7 & 55.5 \\
28 &       &       &100.57 &102.08 & 16 &      &      & 59.7 & 59.4 \\
30 &       &       &112.10 &113.99 & 17 &      &      & 63.9 & 63.4 \\
   &       &       &       &       & 18 &      &      & 68.2 & 67.7 \\
 & $\beta_1$ & $\beta_1$ & $\beta_1$ & $\beta_1$ & 19 & & & 72.7 & 72.0 \\
 0 &  17.3 &  15.7 &  20.6 &  20.6 & 20 &      &      & 77.3 & 76.6 \\
 2 &  18.0 &  16.7 &  21.5 &  21.6 & 21 &      &      & 82.1 & 81.3 \\
 4 &  19.5 &  19.0 &  23.5 &  24.0 & 22 &      &      & 87.0 & 86.1 \\
 6 &  21.9 &  22.6 &       &       & 23 &      &      & 91.9 & 91.0 \\
 8 &  24.6 &  27.4 &       &       & 24 &      &      & 97.0 & 96.1 \\
   &       &       &       &       & 25 &      &      &102.1 &101.3 \\
   &       &       &       &       & 26 &      &      &107.4 &106.6 \\
   &       &       &       &       & 27 &      &      &112.7 &112.0 \\

\hline
\end{tabular}
\end{table}

\newpage

\begin{table*}

\caption{Comparison of experimental data \cite{NDS} (upper line) for several $B(E2)$ ratios of $\gamma$-unstable nuclei
to predictions (lower line) by the Bohr Hamiltonian with $\beta$-dependent mass (with $\delta =\lambda =0$), for the
parameter values shown in Table 1. See subsec. \ref{BE2gam} for further discussion. }

\bigskip

\begin{tabular}{l r@{.}l r@{.}l r@{.}l r@{.}l r@{.}l r@{.}l r@{.}l r@{.}l r@{.}l r@{.}l}

\hline
   \multicolumn{1}{l}{nucl.}
   &\multicolumn{2}{c} {$4_1\to 2_1 \over 2_1\to 0_1$}
    &\multicolumn{2}{c} {$6_1\to 4_1 \over 2_1\to 0_1$}
    &\multicolumn{2}{c} {$8_1\to 6_1 \over 2_1\to 0_1$}
   &\multicolumn{2}{c} {$10_1\to 8_1 \over 2_1\to 0_1$}
    &\multicolumn{2}{c} {$2_2 \to 2_1 \over 2_1\to 0_1$}
   &\multicolumn{2}{c}{$2_2 \to 0_1 \over 2_1\to 0_1$}
   &\multicolumn{2}{c}{$0_2 \to 2_1 \over 2_1\to 0_1$}
   &\multicolumn{2}{c}{$2_3 \to 0_1 \over 2_1 \to 0_1$}  \\

   & \omit\span & \omit\span & \omit\span & \omit\span &
  \omit\span &  \multicolumn{2}{c} {x $10^3$} &  \omit\span &
  \multicolumn{2}{c} {x $10^3$}
   \\

\hline 

$^{98}$Ru   & 1&44(25) & \omit\span & \omit\span & \omit\span &
1&62(61) &  36&0(152)      & \omit\span & \omit\span \\
           & 1&82 & 2&62 & 3&42 & 4&22 & 1&82 & 0&0 & 1&36 & 3&60 \\
           
$^{100}$Ru   & 1&45(13) & \omit\span & \omit\span & \omit\span &
0&64(12) &  41&1(52)      & 0&98(15) & \omit\span \\
           & 1&72 & 2&40 & 3&07 & 3&73 & 1&72 & 0&0 & 1&05 & 10&89 \\

$^{102}$Ru   & 1&50(24) & \omit\span & \omit\span & \omit\span &
0&62(7) &  24&8(7)      & 0&80(14) & \omit\span \\
           & 1&78 & 2&54 & 3&28 & 4&01 & 1&78 & 0&0 & 1&27 & 8&70 \\

$^{104}$Ru   & 1&18(28) & \omit\span & \omit\span & \omit\span &
0&63(15) &  35&0(84)      & 0&42(7) & \omit\span \\
           & 1&63 & 2&18 & 2&71 & 3&21 & 1&63 & 0&0 & 0&79 &22&41 \\

$^{102}$Pd   & 1&56(19) & \omit\span & \omit\span & \omit\span &
0&46(9) &  128&8(735)      & \omit\span & \omit\span \\
           & 1&76 & 2&49 & 3&19 & 3&87 & 1&76 & 0&0 & 1&22 & 12&34 \\
 
$^{104}$Pd   & 1&36(27) & \omit\span & \omit\span & \omit\span &
0&61(8) &  33&3(74)      & \omit\span & \omit\span \\
           & 1&74 & 2&45 & 3&15 & 3&85 & 1&74 & 0&0 & 1&11 & 8&13 \\  
           
 $^{106}$Pd   & 1&63(28) & \omit\span & \omit\span & \omit\span &
0&98(12) &  26&2(31)      & 0&67(18) & \omit\span \\
           & 1&85 & 2&67 & 3&49 & 4&28 & 1&85 & 0&0 & 1&49 & 5&98 \\
           
$^{108}$Pd   & 1&47(20) & 2&16(28) & 2&99(48) & \omit\span &
1&43(14) &  16&6(18)      & 1&05(13) & 1&90(29) \\
           & 1&75 & 2&45 & 3&12 & 3&75 & 1&75 & 0&0 & 1&20 & 15&82 \\ 
           
$^{110}$Pd   & 1&71(34) & \omit\span & \omit\span & \omit\span &
0&98(24) &  14&1(22)      & 0&64(10) & \omit\span \\
           & 1&76 & 2&43 & 3&01 & 3&51 & 1&76 & 0&0 & 1&31 & 26&24 \\ 
           
$^{106}$Cd   & 1&78(25) & \omit\span & \omit\span & \omit\span &
0&43(12) &  93&0(127)      & \omit\span & \omit\span \\
           & 1&68 & 2&32 & 2&95 & 3&58 & 1&68 & 0&0 & 0&92 & 10&44 \\
           
$^{108}$Cd   & 1&54(24) & \omit\span & \omit\span & \omit\span &
0&64(20) &  67&7(120)      & \omit\span & \omit\span \\
           & 1&85 & 2&69 & 3&52 & 4&35 & 1&85 & 0&0 & 1&49 & 4&06 \\
           
$^{110}$Cd   & 1&68(24) & \omit\span & \omit\span & \omit\span &
1&09(19) &  48&9(78)      & \omit\span & 9&85(595) \\
           & 1&99 & 2&97 & 3&93 & 4&87 & 1&99 & 0&0 & 1&98 & 1&61 \\
           
$^{112}$Cd   & 2&02(22) & \omit\span & \omit\span & \omit\span &
0&50(10) &  19&9(35)      & 1&69(48) & 11&26(210) \\
           & 2&00 & 2&99 & 3&98 & 4&96 & 2&00 & 0&0 & 1&99 & 0&48 \\ 
           
$^{114}$Cd   & 1&99(25) & 3&83(72) & 2&73(97) & \omit\span &
0&71(24) &  15&4(29)      & 0&88(11) & 10&61(193) \\
           & 2&00 & 2&99 & 3&97 & 4&94 & 2&00 & 0&0 & 1&99 & 0&74 \\   
           
$^{116}$Cd   & 1&70(52) & \omit\span & \omit\span & \omit\span &
0&63(46) &  32&8(86)      & 0&02 & \omit\span \\
           & 1&74 & 2&46 & 3&17 & 3&90 & 1&74 & 0&0 & 1&11 & 4&42 \\  
           
$^{118}$Cd   & $>$1&85 & \omit\span & \omit\span & \omit\span &
\omit\span & \omit\span      & 0&16(4) & \omit\span \\
           & 1&71 & 2&39 & 3&06 & 3&74 & 1&71 & 0&0 & 1&00 & 5&88 \\  
           
$^{118}$Xe   & 1&11(7) & 0&88(27) & 0&49(20) & $>$0&73 &
\omit\span & \omit\span & \omit\span &\omit\span \\
           & 1&67 & 2&28 & 2&85 & 3&39 & 1&67 & 0&0 & 0&95 & 21&93 \\  
 
$^{120}$Xe   & 1&16(14) & 1&17(24) & 0&96(22) & 0&91(19) &
\omit\span & \omit\span & \omit\span &\omit\span \\
           & 1&60 & 2&11 & 2&60 & 3&08 & 1&60 & 0&0 & 0&67 & 21&51 \\            

$^{122}$Xe   & 1&47(38) & 0&89(26) & $>$0&44 & \omit\span &
   \omit\span & \omit\span    & \omit\span & \omit\span \\
           & 1&58 & 2&05 & 2&48 & 2&89 & 1&58 & 0&0 & 0&63 & 29&29 \\
           
$^{124}$Xe   & 1&34(24) & 1&59(71) & 0&63(29) & 0&29(8) &
0&70(19) &  15&9(46)      & \omit\span  & \omit\span \\
           & 1&59 & 2&09 & 2&57 & 3&04 & 1&59 & 0&0 & 0&63 & 20&14 \\
           
$^{128}$Xe   & 1&47(20) & 1&94(26) & 2&39(40) & 2&74(114) &
1&19(19) &  15&9(23)      & \omit\span  & \omit\span \\
           & 1&63 & 2&20 & 2&75 & 3&31 & 1&63 & 0&0 & 0&73 & 9&64 \\
  
$^{132}$Xe   & 1&24(18) & \omit\span & \omit\span & \omit\span &
1&77(29) &  3&4(7)      & \omit\span & \omit\span \\
           & 2&00 & 3&00 & 4&00 & 5&00 & 2&00 & 0&0 & 2&00 & 0&00 \\  
           
$^{130}$Ba   & 1&36(6) & 1&62(15) & 1&55(56) & 0&93(15) &
\omit\span  & \omit\span   & \omit\span  & \omit\span \\
           & 1&56 & 2&01 & 2&41 & 2&77 & 1&56 & 0&0 & 0&61 & 34&54 \\
           
$^{132}$Ba   & \omit\span & \omit\span & \omit\span & \omit\span &
3&35(64) &  90&7(177)      & \omit\span & \omit\span \\
           & 1&68 & 2&30 & 2&90 & 3&50 & 1&68 & 0&0 & 0&92 & 15&21 \\  
  
$^{134}$Ba   & 1&55(21) & \omit\span & \omit\span & \omit\span &
2&17(69) &  12&5(41)      & \omit\span & \omit\span \\
           & 1&75 & 2&48 & 3&21 & 3&94 & 1&75 & 0&0 & 1&14 & 4&08 \\    
  
$^{142}$Ba   & 1&40(17) & 0&56(14) & \omit\span & \omit\span &
\omit\span & \omit\span   & \omit\span & \omit\span \\
           & 1&55 & 2&00 & 2&41 & 2&82 & 1&55 & 0&0 & 0&49 & 18&60 \\ 
             
\hline
\end{tabular}
\end{table*}  
  
\begin{table*}

\setcounter{table}{3} \caption{ (continued) }

\bigskip

\begin{tabular}{l r@{.}l r@{.}l r@{.}l r@{.}l r@{.}l r@{.}l r@{.}l r@{.}l r@{.}l r@{.}l}

\hline
   \multicolumn{1}{l}{nucl.}
   &\multicolumn{2}{c} {$4_1\to 2_1 \over 2_1\to 0_1$}
    &\multicolumn{2}{c} {$6_1\to 4_1 \over 2_1\to 0_1$}
    &\multicolumn{2}{c} {$8_1\to 6_1 \over 2_1\to 0_1$}
   &\multicolumn{2}{c} {$10_1\to 8_1 \over 2_1\to 0_1$}
    &\multicolumn{2}{c} {$2_2 \to 2_1 \over 2_1\to 0_1$}
   &\multicolumn{2}{c}{$2_2 \to 0_1 \over 2_1\to 0_1$}
   &\multicolumn{2}{c}{$0_2 \to 2_1 \over 2_1\to 0_1$}
   &\multicolumn{2}{c}{$2_3 \to 0_1 \over 2_1 \to 0_1$}  \\

   & \omit\span & \omit\span & \omit\span & \omit\span &
  \omit\span &  \multicolumn{2}{c} {x $10^3$} &  \omit\span &
  \multicolumn{2}{c} {x $10^3$}
   \\
           
\hline 

$^{148}$Nd   & 1&61(13) & 1&76(19) & \omit\span & \omit\span &
0&25(4) & 9&3(17)   & 0&54(6) & 32&82(816) \\
           & 1&63 & 2&17 & 2&68 & 3&15 & 1&63 & 0&0 & 0&81 & 26&86 \\ 
           
$^{152}$Gd   & 1&84(29) & 2&74(81) & \omit\span & \omit\span &
0&23(4) & 4&2(8)   & 2&47(78) & \omit\span \\
           & 1&98 & 2&92 & 3&81 & 4&65 & 1&98 & 0&0 & 1&95 & 4&51 \\
  
$^{154}$Dy   & 1&62(35) & 2&05(42) & 2&27(62) & 1&86(69) &
\omit\span   & \omit\span & \omit\span & \omit\span \\
           & 1&91 & 2&79 & 3&64 & 4&46 & 1&91 & 0&0 & 1&70 & 5&41 \\
  
$^{156}$Er   & 1&78(16) & 1&89(36) & 0&76(20) & 0&88(22) &
\omit\span   & \omit\span & \omit\span & \omit\span \\
           & 1&70 & 2&35 & 3&00 & 3&64 & 1&70 & 0&0 & 0&98 & 11&50 \\
  
$^{192}$Pt   & 1&56(12) & 1&23(55) & \omit\span & \omit\span &
1&91(16)   & 9&5(9) & \omit\span & \omit\span \\
           & 1&59 & 2&09 & 2&57 & 3&05 & 1&59 & 0&0 & 0&61 & 16&98 \\
            
 $^{194}$Pt   & 1&73(13) & 1&36(45) & 1&02(30) & 0&69(19) &
1&81(25) &  5&9(9)      & 0&01  & \omit\span \\
           & 1&59 & 2&09 & 2&57 & 3&04 & 1&59 & 0&0 & 0&63 & 19&78 \\
           
 $^{196}$Pt   & 1&48(3) & 1&80(23) & 1&92(23) & \omit\span &
\omit\span &  0&4      & 0&07(4)  & 0&06(6) \\
           & 1&64 & 2&21 & 2&75 & 3&28 & 1&64 & 0&0 & 0&82 & 20&83 \\ 
           
 $^{198}$Pt   & 1&19(13) & $>$1&78 & \omit\span & \omit\span &
 1&16(23) &  1&2(4)      & 0&81(22)  & 1&56(126) \\
           & 1&82 & 2&60 & 3&36 & 4&08 & 1&82 & 0&0 & 1&41 & 10&09 \\          
           
\hline
\end{tabular}
\end{table*}

\begin{table*}

\caption{Comparison of experimental data \cite{NDS} (upper line) for several $B(E2)$ ratios of axially symmetric prolate deformed nuclei
to predictions (lower line) by the Bohr Hamiltonian with $\beta$-dependent mass (with $\delta =\lambda =0$), for the
parameter values shown in Table 2. See subsec. \ref{BE2def} for further discussion. }

\bigskip

\begin{tabular}{l r@{.}l r@{.}l r@{.}l r@{.}l r@{.}l r@{.}l r@{.}l r@{.}l r@{.}l r@{.}l}

\hline
   \multicolumn{1}{l}{nucl.}
   &\multicolumn{2}{c} {$4_1\to 2_1 \over 2_1\to 0_1$}
    &\multicolumn{2}{c} {$6_1\to 4_1 \over 2_1\to 0_1$}
    &\multicolumn{2}{c} {$8_1\to 6_1 \over 2_1\to 0_1$}
   &\multicolumn{2}{c} {$10_1\to 8_1 \over 2_1\to 0_1$}
    &\multicolumn{2}{c} {$2_\beta \to 0_1 \over 2_1\to 0_1$}
   &\multicolumn{2}{c}{$2_\beta \to 2_1 \over 2_1\to 0_1$}
   &\multicolumn{2}{c}{$2_\beta \to 4_1 \over 2_1\to 0_1$}
   &\multicolumn{2}{c}{$2_\gamma\to 0_1 \over 2_1 \to 0_1$}
   &\multicolumn{2}{c}{$2_\gamma\to 2_1 \over 2_1 \to 0_1$} 
   &\multicolumn{2}{c}{$2_\gamma\to 4_1 \over 2_1 \to 0_1$} \\

   & \omit\span & \omit\span & \omit\span & \omit\span &
  \multicolumn{2}{c} {x $10^3$} &  \multicolumn{2}{c} {x $10^3$} &  \multicolumn{2}{c} {x $10^3$} &
  \multicolumn{2}{c} {x $10^3$} &  \multicolumn{2}{c} {x $10^3$} &  \multicolumn{2}{c} {x $10^3$}
   \\

\hline $^{154}$Sm   & 1&40(5) & 1&67(7) & 1&83(11) & 1&81(11) &
5&4(13) &  \omit\span      & \multicolumn{2}{c} {25(6)} &
18&4(34) &  \omit\span & 3&9(7) \\
           & 1&47 & 1&69 & 1&87 & 2&06 & 26&7 & 50&0 &  \multicolumn{2}{c}  {150} &
47&5 & 69&6 & 3&7 \\

$^{156}$Gd & 1&41(5) & 1&58(6) & 1&71(10) & 1&68(9) &  3&4(3) &
\multicolumn{2}{c} {18(2)} & \multicolumn{2}{c} {22(2)} &
25&0(15) & 38&7(24) & 4&1(3)  \\
           & 1&48 & 1&73 & 1&95 & 2&18 & 29&7 & 59&1 & \multicolumn{2}{c} {191} &
62&5 & 92&4  & 4&9  \\

$^{158}$Gd & 1&46(5) &  \omit\span        & 1&67(16) & 1&72(16) &
1&6(2) & 0&4(1) & 7&0(8) &
17&2(20) & 30&3(45) & 1&4(2) \\
           & 1&46 & 1&66 & 1&82 & 1&98 & 25&7 & 45&9 & \multicolumn{2}{c} {127} &
64&0 & 93&0 &  4&8 \\

$^{158}$Dy & 1&45(10) & 1&86(12) & 1&86(38) & 1&75(28) &
\multicolumn{2}{c} {12(3)} & \multicolumn{2}{l} {19(4)} &
\multicolumn{2}{c} {66(16)} &
32&2(78) & 103&8(258) & 11&5(48) \\
           & 1&50 & 1&78 & 2&04 & 2&31 & 30&5 & 65&4 & \multicolumn{2}{c} {232} &
88&5 & 131&7  &  7&1 \\

$^{160}$Dy & 1&46(7) & 1&23(7) & 1&70(16) & 1&69(9) &  3&4(4) &
\omit\span   & 8&5(10) &
23&2(21) & 43&8(42) & 3&1(3) \\
           & 1&46 & 1&68 & 1&85 & 2&03 & 22&9 & 43&5 & \multicolumn{2}{c} {133} &
78&6 & 114&5 & 6&0 \\

$^{162}$Dy & 1&45(7) & 1&51(10) & 1&74(10) & 1&76(13) & \omit\span
&  \omit\span   &  \omit\span   &
0&12(1) & 0&20  & 0&02\\
           & 1&45 & 1&65 & 1&80 & 1&95 & 23&9 & 42&4 & \multicolumn{2}{c} {116} &
89&8 & 129&8 & 6&7 \\

$^{164}$Dy & 1&30(7) & 1&56(7) & 1&48(9) & 1&69(9) &  \omit\span &
\omit\span &  \omit\span   &
19&1(22) & 38&3(39) & 4&6(5) \\
           & 1&44 & 1&62 & 1&75 & 1&86 & 16&9 & 29&1 & \multicolumn{2}{c} {77} &
99&7 & 143&4  & 7&3  \\

$^{162}$Er &  \omit\span         &  \omit\span                &
\omit\span  & \omit\span     & \multicolumn{2}{c} {8(7)} &
\omit\span  & \multicolumn{2}{c} {170(90)} &
32&5(28) & 77&0(56) &  9&4(69) \\
           & 1&49 & 1&75 & 1&99 & 2&24 & 27&8 & 58&3 & \multicolumn{2}{c} {202} &
91&1 & 134&8 & 7&2 \\

$^{164}$Er & 1&18(13) &   \omit\span        & 1&57(9) & 1&64(11) &
\omit\span &  \omit\span  &  \omit\span   &
23&9(35) & 52&3(72) & 7&8(12) \\
           & 1&47 & 1&70 & 1&89 & 2&09 & 28&3 & 53&5 & \multicolumn{2}{c} {162} &
103&8 & 151&2 & 7&9 \\

$^{166}$Er & 1&45(12) & 1&62(22) & 1&71(25) & 1&73(23) &
\omit\span    &  \omit\span   &  \omit\span   &
25&7(31) & 45&3(54) & 3&1(4) \\
           & 1&46 & 1&66 & 1&81 & 1&96 & 20&7 & 38&2 & \multicolumn{2}{c} {111} &
100&0 & 144&8  & 7&4    \\

$^{168}$Er & 1&54(7) & 2&13(16) & 1&69(11) & 1&46(11) & \omit\span
&  \omit\span   &  \omit\span   &
23&2(15) & 41&1(31) & 3&0(3) \\
           & 1&45 & 1&65 & 1&79 & 1&93 & 27&7 & 47&2 & \multicolumn{2}{c} {120} &
100&6 & 145&1 &  7&4  \\

$^{170}$Er &  \omit\span         &  \omit\span         & 1&78(15)
& 1&54(11) & 1&4(1) & 0&2(2) & 6&8(12) &
17&7(9) &  \omit\span        & 1&4(4) \\
           & 1&47 & 1&69 & 1&86 & 2&03 & 39&2 & 67&9 & \multicolumn{2}{c} {177} &
78&6 & 114&2 & 5&9  \\

$^{166}$Yb & 1&43(9) & 1&53(10) & 1&70(18) & 1&61(80) & \omit\span
&  \omit\span   &  \omit\span   &
     \omit\span    &  \omit\span    &  \omit\span        \\
           & 1&50 & 1&78 & 2&05 & 2&33 &  33&7 & 71&0 & \multicolumn{2}{c} {245} &
97&2 & 144&5 & 7&8 \\

$^{168}$Yb &  \omit\span         &   \omit\span             &
\omit\span  & \omit\span       & 8&6(9) &  \omit\span   &
\omit\span   &
22&0(55) & 45&9(73) & 8&6 \\
           & 1&48 & 1&72 & 1&93 & 2&14 & 29&6 & 57&5 & \multicolumn{2}{c} {180} &
82&9 & 121&6 & 6&4  \\

$^{170}$Yb &  \omit\span         &  \omit\span        & 1&79(16) &
1&77(14) & 5&4(10) &  \omit\span   &  \omit\span   &
13&4(34) & 23&9(57) & 2&4(6) \\
           & 1&47 & 1&71 & 1&91 & 2&12 & 30&6 & 58&2 & \multicolumn{2}{c} {176} &
66&2 & 97&1 & 5&1  \\

$^{172}$Yb & 1&42(10) & 1&51(14) & 1&89(19) & 1&77(11) & 1&1(1) &
3&7(6) & \multicolumn{2}{c} {12(1)} & 6&3(6) & \omit\span   & 0&6(1) \\
           & 1&46 & 1&67 & 1&83 & 1&99 & 32&2 & 55&9 & \multicolumn{2}{c} {147} &
51&6 & 75&0 & 3&9  \\

$^{174}$Yb & 1&39(7) & 1&84(26) & 1&93(12) & 1&67(12) & \omit\span
&  \omit\span  &  \omit\span  &
    \omit\span  & 12&4(29)   & \omit\span          \\
           & 1&45 & 1&63 & 1&75 & 1&86 & 20&9 & 35&1 & \multicolumn{2}{c} {88} &
45&0 & 64&9 & 3&3  \\

$^{176}$Yb & 1&49(15) & 1&63(14) & 1&65(28) & 1&76(18) &
\omit\span    &  \omit\span   &  \omit\span   &
9&8
 &  \omit\span   &  \omit\span       \\
           & 1&46 & 1&66 & 1&82 & 1&97 & 27&9 & 49&0 & \multicolumn{2}{c} {132} &
63&1 & 91&6 & 4&7  \\

$^{174}$Hf &  \omit\span            &   \omit\span       &
\omit\span &  \omit\span  & \multicolumn{2}{c}{14(4)} & \omit\span
& \multicolumn{2}{c} {9(3)} &
31&6(161) & 48&7(124) & \omit\span        \\
           & 1&48 & 1&74 & 1&96 & 2&20 & 31&4 & 62&2 & \multicolumn{2}{c}{200} &
66&9 & 98&8 & 5&3  \\

$^{176}$Hf & \omit\span  & \omit\span      & \omit\span       &
\omit\span & 5&4(11) & \omit\span & \multicolumn{2}{c} {31(6)} &
21&3(26) & \omit\span  &   \omit\span     \\
           & 1&47 & 1&70 & 1&89 & 2&09 & 30&8 & 57&3 & \multicolumn{2}{c} {169} &
57&9 & 84&9 & 4&5 \\

$^{178}$Hf & \omit\span       & 1&38(9) & 1&49(6) & 1&62(7) &
0&4(2) & \omit\span & 2&4(9) &
24&5(39) & 27&7(28) & 1&6(2) \\
           & 1&47 & 1&69 & 1&88 & 2&07 & 28&4 & 53&1 & \multicolumn{2}{c} {158} &
73&8 & 107&8 & 5&6 \\

$^{180}$Hf & 1&48(20) & 1&41(15) & 1&61(26) & 1&55(10) &
\omit\span & \omit\span & \omit\span &
24&5(47) & 32&9(56)  & \omit\span        \\
           & 1&46 & 1&66 & 1&82 & 1&98 & 34&9 & 59&5 & \multicolumn{2}{c} {151} &
78&4 & 113&4 & 5&8 \\

$^{182}$W  & 1&43(8) & 1&46(16) & 1&53(14) & 1&48(14) &  6&6(6) &
4&6(6) & \multicolumn{2}{c} {13(1)} &
24&8(12) & 49&2(24) & 0&2\\
           & 1&47 & 1&69 & 1&87 & 2&04 & 32&5 & 58&3 & \multicolumn{2}{c} {162} &
79&9 & 116&2 & 6&0  \\

$^{184}$W  & 1&35(12) & 1&54(9) & 2&00(18) & 2&45(51) & 1&8(3) &
\omit\span & \multicolumn{2}{c} {24(3)} &
37&1(28) & 70&6(51) & 4&0(4) \\
           & 1&48 & 1&73 & 1&95 & 2&16 &  40&7 & 75&2 & \multicolumn{2}{c} {216} &
128&3 & 187&3 & 9&8 \\

$^{186}$W  & 1&30(9) & 1&69(12) & 1&60(12) & 1&36(36) & \omit\span & \omit\span
& \omit\span & 41&7(92) & 91&0(201) & \omit\span        \\
           & 1&51 & 1&80 & 2&07 & 2&34 & 46&2 & 91&9 & \multicolumn{2}{c} {289} &
165&7 & 244&5 & 12&9 \\

$^{186}$Os & 1&45(7) & 1&99(7) & 1&89(11) & 2&06(44) & \omit\span
& \omit\span & \omit\span &
109&4(71) & 254&6(150) & 13&0(47) \\
           & 1&53 & 1&87 & 2&20 & 2&55 & 39&7 & 90&2 & \multicolumn{2}{c} {335} &
164&9 & 247&4 & 13&4 \\

$^{188}$Os & 1&68(11) & 1&75(11) & 2&04(15) & 2&38(32)        &
\omit\span & \omit\span & \omit\span &
63&3(92) & 202&5(304) & 43&0(74) \\
           & 1&54 & 1&89 & 2&25 & 2&63 & 33&9 & 83&9 & \multicolumn{2}{c} {344} &
229&8 & 345&2 & 18&7 \\

\hline
\end{tabular}
\end{table*}

\begin{table*}
\setcounter{table}{4} \caption{ (continued) }

\bigskip

\begin{tabular}{l r@{.}l r@{.}l r@{.}l r@{.}l r@{.}l r@{.}l r@{.}l r@{.}l r@{.}l r@{.}l}

\hline
   \multicolumn{1}{l}{nucl.}
   &\multicolumn{2}{c} {$4_1\to 2_1 \over 2_1\to 0_1$}
    &\multicolumn{2}{c} {$6_1\to 4_1 \over 2_1\to 0_1$}
    &\multicolumn{2}{c} {$8_1\to 6_1 \over 2_1\to 0_1$}
   &\multicolumn{2}{c} {$10_1\to 8_1 \over 2_1\to 0_1$}
    &\multicolumn{2}{c} {$2_\beta \to 0_1 \over 2_1\to 0_1$}
   &\multicolumn{2}{c}{$2_\beta \to 2_1 \over 2_1\to 0_1$}
   &\multicolumn{2}{c}{$2_\beta \to 4_1 \over 2_1\to 0_1$}
   &\multicolumn{2}{c}{$2_\gamma\to 0_1 \over 2_1 \to 0_1$}
   &\multicolumn{2}{c}{$2_\gamma\to 2_1 \over 2_1 \to 0_1$} 
   &\multicolumn{2}{c}{$2_\gamma\to 4_1 \over 2_1 \to 0_1$} \\

   & \omit\span & \omit\span & \omit\span & \omit\span &
  \multicolumn{2}{c} {x $10^3$} &  \multicolumn{2}{c} {x $10^3$} &  \multicolumn{2}{c} {x $10^3$}
   & \multicolumn{2}{c} {x $10^3$} &  \multicolumn{2}{c} {x $10^3$} &  \multicolumn{2}{c} {x $10^3$}
   \\
   
\hline 

$^{230}$Th & 1&36(8)         & \omit\span      & \omit\span &
\omit\span & 5&7(26) & \omit\span & \multicolumn{2}{c} {20(11)} &
15&6(59) & 28&1(100) & 1&8(11) \\
           & 1&47 & 1&70 & 1&90 & 2&09  & 30&0 & 56&4 & \multicolumn{2}{c} {168} &
63&6 & 93&2 & 4&9 \\

$^{232}$Th & 1&44(15) & 1&65(14) & 1&73(12) & 1&82(15) &
\multicolumn{2}{c}{14(6)} & 2&6(13) & \multicolumn{2}{c} {17(8)} &
14&6(28) & 36&4(56) & 0&7 \\
           & 1&46 & 1&67 & 1&84 & 2&01 & 25&8 & 47&1 & \multicolumn{2}{c}{135} &
57&0 & 83&0 & 4&3 \\

$^{234}$U  & \omit\span &   \omit\span     & \omit\span
&\omit\span &\omit\span &\omit\span & \omit\span &
12&5(27) & 21&1(44) & 1&2(3) \\
           & 1&45 & 1&64 & 1&78 & 1&90 & 20&7 & 36&1 & \multicolumn{2}{c}{97} &
42&7 & 61&8 & 3&2  \\

$^{236}$U  & 1&42(11) & 1&55(11) & 1&59(17) & 1&46(17) &
\omit\span & \omit\span & \omit\span & \omit\span
      &   \omit\span    &   \omit\span    \\
           & 1&45 & 1&63 & 1&76 & 1&87 & 19&3 & 33&2 & \multicolumn{2}{c}{87} &
44&7 & 64&5 & 3&3 \\

$^{238}$U  & \omit\span       & \omit\span       & 1&45(23) &
1&71(22) & 1&4(6) & 3&6(14) & \multicolumn{2}{c}{12(5)} &
10&8(8) & 18&9(17)  & 1&2(1) \\
           & 1&45 & 1&63 & 1&75 & 1&86 & 18&9 & 32&3 & \multicolumn{2}{c}{83} &
37&7 & 54&5 & 2&8 \\

$^{238}$Pu &  \omit\span            &  \omit\span     & \omit\span
& \omit\span& \multicolumn{2}{c}{14(4)} &\omit\span &
\multicolumn{2}{c}{11(4)} &
   \omit\span   &  \omit\span   &  \omit\span      \\
           & 1&44 & 1&62 & 1&73 & 1&84 & 19&1 & 31&7 &  \multicolumn{2}{c}{78} &
41&6 & 59&9 & 3&0 \\

$^{250}$Cf &  \omit\span            &  \omit\span     & \omit\span
&\omit\span & \omit\span&\omit\span &\omit\span &
6&8(17) & 10&9(25) & 0&6(1) \\
           & 1&44 & 1&61 & 1&72 & 1&81 & 15&0 & 24&9 & \multicolumn{2}{c}{61} &
40&0 & 57&5 & 2&9  \\
\hline
\end{tabular}
\end{table*}

\end{document}